\documentclass[lettersize,journal]{IEEEtran}
\usepackage{amsmath,amssymb,amsfonts}
\usepackage{algorithm}
\usepackage{algpseudocode}
\usepackage{algorithm}
\usepackage{array}
\usepackage[caption=false,font=normalsize,labelfont=sf,textfont=sf]{subfig}
\usepackage{textcomp}
\usepackage{stfloats}
\usepackage{url}
\usepackage{verbatim}
\usepackage{graphicx}
\usepackage{booktabs}
\usepackage{multirow}
\usepackage{cite}
\usepackage{hyperref}
\hypersetup{
    pdfborder={0 0 0}  % 去除链接边框
}
\usepackage{xcolor}
\begin{document}

\title{Diffusion-driven SpatioTemporal \\Graph KANsformer for \\Medical Examination Recommendation
}

\author{Jianan Li, Yangtao Zhou, Zhifu Zhao*, Qinglan Huang, Jian Qi, Xiao He, Hua Chu, Fu Li
        % <-this % stops a space
\thanks{This work was supported by the National Natural Science Foundation of China under Grant No.62302373, No.62176201, and No.62202356, the Key Research and Development Program of Shaanxi under Grant No.2023-ZDLSF-07, the STI 2030—Major Projects under Grant No.2022ZD0208500, and the Fundamental Research Funds of the Central Universities of China under Grants No.ZYTS24092, NO.OTZX24085, No.ZYTS23065, No.QTZX2107, No.RW200141, and No.2242023k30022. (Jianan Li and Yangtao Zhou are co-first authors) (Corresponding authors: Zhifu Zhao.)}
\thanks{Jianan Li, Yangtao Zhou, Jian Qi, Xiao He, and Hua Chu are with the School of Computer Science and Technology, Xidian University, Xi'an 710071, China (e-mail: lijianan@xidian.edu.cn; zhou\_yt@stu.xidian.edu.cn; jianqi@stu.xidian.edu.cn; xavierho@stu.xidian.edu.cn; hchu@mail.xidian.edu.cn).

Zhifu Zhao, Qinglan Huang, and Fu Li are with the School of Artificial Intelligence, Xidian University, Xi'an 710071, China (e-mail: zfzhao@xidian.edu.cn; qlhuang@stu.xidian.edu.cn; fuli@mail.xidian.edu.cn)}
}

\maketitle

\markboth{Journal of \LaTeX\ Class Files,~Vol.~14, No.~8, August~2021}%
{Shell \MakeLowercase{\textit{et al.}}: A Sample Article Using IEEEtran.cls for IEEE Journals}

%\thanks{This paper was produced by the IEEE Publication Technology Group. They are in Piscataway, NJ.}% <-this % stops a space
%\thanks{Manuscript received April 19, 2021; revised August 16, 2021.}


% 在第二列中清除 IEEEpubid 标记
%\IEEEpubidadjcol


%\maketitle
%\vspace{-10mm}  % 调整标题部分与后续内容之间的空白

\begin{abstract}

Recommendation systems in AI-based medical diagnostics and treatment constitute a critical component of AI in healthcare. Although some studies have explored this area and made notable progress, healthcare recommendation systems remain in their nascent stage. And these researches mainly target the treatment process such as drug or disease recommendations. In addition to the treatment process, the diagnostic process, particularly determining which medical examinations are necessary to evaluate the condition, also urgently requires intelligent decision support. To bridge this gap, we first formalize the task of medical examination recommendations. 
Compared to traditional recommendations, the medical examination recommendation involves more complex interactions. This complexity arises from two folds: 1) The historical medical records for examination recommendations are heterogeneous and redundant, which makes the recommendation results susceptible to noise. 2) The correlation between the medical history of patients is often irregular, making it challenging to model spatiotemporal dependencies. Motivated by the above observation, we propose a novel Diffusion-driven SpatioTemporal Graph KANsformer for Medical Examination Recommendation (DST-GKAN) with a two-stage learning paradigm to solve the above challenges. In the first stage, we exploit a task-adaptive diffusion model to distill recommendation-oriented information by reducing the noises in heterogeneous medical data. In the second stage, a spatiotemporal graph KANsformer is proposed to simultaneously model the complex spatial and temporal relationships. Moreover, to facilitate the medical examination recommendation research and verify the effectiveness of our proposed method, we introduce a comprehensive dataset. The experimental results demonstrate the state-of-the-art performance of the proposed method compared to various competitive baselines.
\end{abstract}


\section{Introduction}

\IEEEPARstart{H}{ealth} care is the maintenance of health through the prevention, diagnosis, and treatment, which is a key concern in most countries. Recently, considering the aging population and the ever-increasing number of patients with chronic diseases, the imbalance between the supply and demand of medical resources is becoming increasingly severe. With the widespread adoption of Electronic Health Records (EHRs) and the evolution of Artificial Intelligence (AI) technology, AI-based medical diagnostic and treatment systems (AI-MDTS) have emerged as promising approaches that leverage data-driven decision support to alleviate the strain on medical resources. And the recommender systems \cite{lopez2012property} that aims to provide decision support for doctors and patients by analyzing historical medical records sequence and current health assessments, is a vital aspect of these AI-based approaches.

\begin{figure}[t]
    \centering
    \includegraphics[width=1\linewidth]{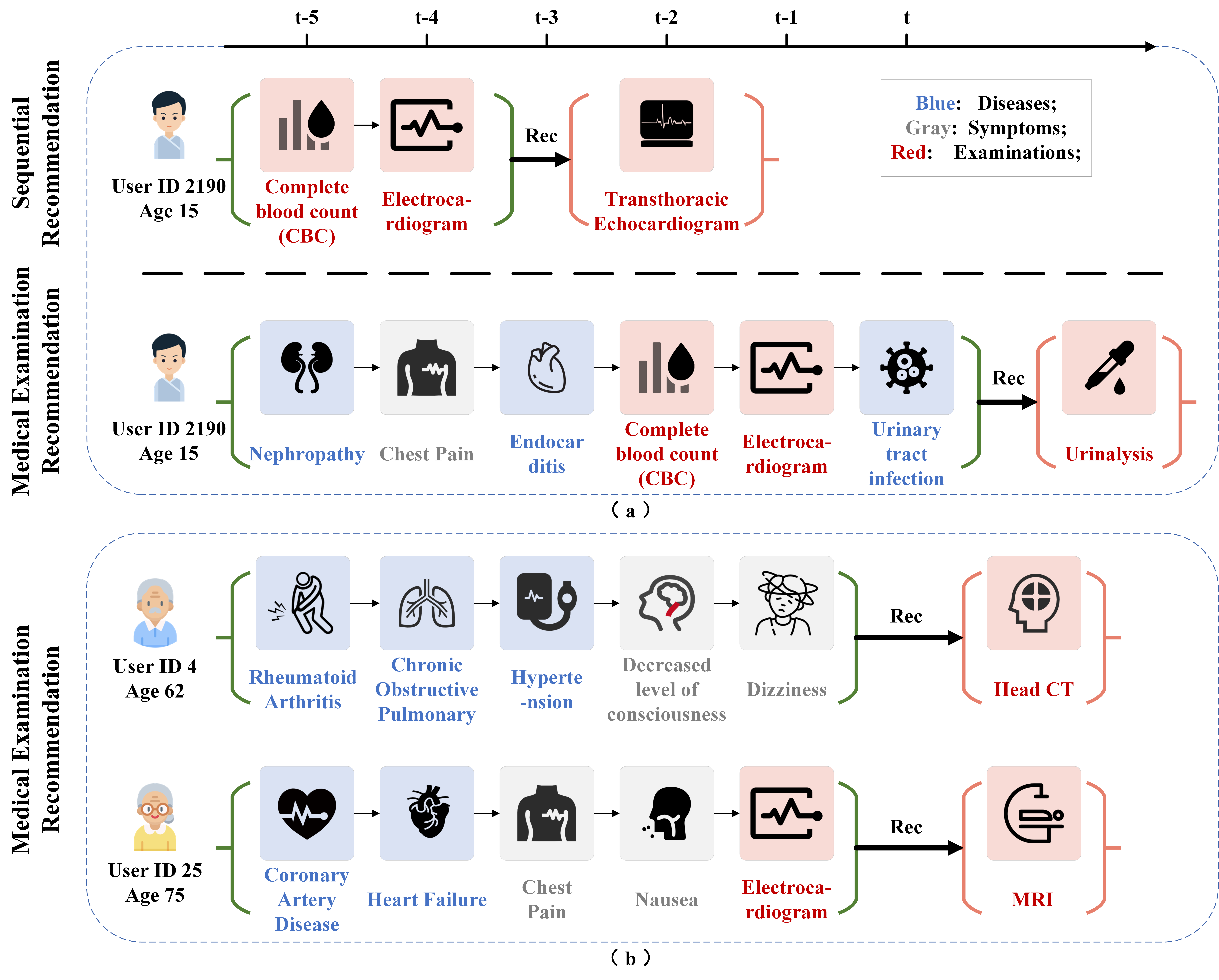}
    \vspace{-0.5cm}
    \caption{Toy examples of medical examination recommendation. (a) comparison among medical examination recommendation and existing sequential recommendation methods. (b) Medical examination recommendation with heterogeneity property. Blue indicates diseases, gray represents symptoms red highlights the recommended examinations.}
    \label{fig:1 introduction}
\end{figure}
%\vspace{-0.8cm}

To achieve intelligent medical recommendations, a number of deep learning based sequential recommendation models are proposed for medical recommendation. Shang et al \cite{shang2019gamenet} proposed GAMENet to integrate drug interaction knowledge through GCN-based modules and model patient records as an EHR graph for drug recommendation. Baytas et al. \cite{baytas2017patient} introduced T-LSTM to handle irregular time intervals in longitudinal patient records for patient subtypes. Choi et al. \cite{choi2018mime} developed MiME to convert EHR data into multilevel embeddings, overcoming data scarcity challenges for predictive healthcare. Ma et al. \cite{ma2018health} designed Health-ATM by combining time-aware Convolutional Neural Network (CNN) and attention mechanisms, providing comprehensive health insights for risk prediction.
%这里添加几篇医疗的参考文献。
Despite their significant advancements and recent popularity, the healthcare recommendation systems are still in their infancy. Specifically, these methods mainly focus on the treatment process (eg. drugs or diseases) in healthcare. In addition to the treatment process, the diagnostic process, particularly determining which medical examinations are necessary to evaluate the condition, also significant for intelligent decision support. However, the medical examination recommendation task remains unexplored.
 
To bridge this gap, in this paper, we first formalize the medical examination recommendations task with the goal of providing intelligent decision support to determine the appropriate examinations. Compared to traditional user-item recommendations, the medical examination recommendations involves more complex interactions, as shown in Fig \ref{fig:1 introduction}. This complexity arises from two folds: 1) The historical medical records influencing examination recommendations are heterogeneous and redundant, which makes the results susceptible to noise. 2) The correlation between the medical history of patients is often irregular, making it challenging to model sequential dependencies.

To address these challenges, we meticulously design a Diffusion-driven SpatioTemporal Graph KANsformer (DST-GKAN) for medical examination recommendation with a two-stage learning paradigm. Specifically, in the first stage, driven by the inherent applicability of diffusion models \cite{ho2020denoising, sohl2015deep} for denoising tasks, we propose a task-adaptive diffusion model to mitigate the impact of task-irrelative noise in heterogeneous medical data. In the second stage, to effectively capture the spatiotemporal characteristics of complex heterogeneity data, a spatiotemporal graph KANsformer is proposed. In particular, we innovatively integrate the Kolmogorov-Arnold Networks (KANs) \cite{liu2024kan} into the Transformer framework, naming it KANsformer, to enhance the model expressiveness. Moreover, to support the medical examination recommendation research, we introduce a comprehensive dataset. Through extensive experiments using this real-world medical dataset, we demonstrate the superiority of our proposed method compared to various advanced baselines.

The main contribution can be concluded as follows:
\begin{itemize}
\item A meticulously designed Diffusion-driven SpatioTemporal Graph KANsformer model (DST-GKAN) is proposed for Medical Examination Recommendation. To the best of our knowledge, this is the first work to investigate the Medical Examination Recommendation task.
\item We propose a task-adaptive diffusion model to distill recommendation-oriented information by reducing the noises in heterogeneous medical data. Notably, this diffusion model can receive feedback from the recommendation stage for adaptive distillation.
\item To tackle entity heterogeneity and relational complexity, a spatiotemporal graph KANsformer with progressive spatial-temporal aggregation strategy is proposed to simultaneously model complex spatial and temporal relationships. 
\item To mitigate the issues of incompleteness or lack of existing medical examination data, we propose a benchmark dataset with heterogeneity and temporality for Medical Examination recommendation, named MeExam.
\end{itemize}

\section{Related Work}
\subsection{Medical Recommendation}
In recent years, with the increasing adoption of EHRs and the advancement of deep learning technologies, medical recommendation has attracted a great amount of research attention\cite{su2020gate}. Common medical recommendations include drug recommendations, clinical decision-making, tag recommendations and doctor recommendations.

In the domain of drug recommendations, Shang et al. \cite{shang2019gamenet} categorize existing approaches into two primary types: instance-based \cite{Zhang2017leap, tan20224sdrug, liu2024leader} and longitudinal \cite{choi2016retain, su2022tahdnet, liu2022multi, liang2024dual, shang2019gamenet}. Instance-based methods, such as LEAP \cite{Zhang2017leap} and 4SDrug \cite{tan20224sdrug}, provide drug suggestions by focusing on a patient’s immediate health status, whereas longitudinal methods, including RETAIN \cite{choi2016retain} and TAHDNet \cite{su2022tahdnet}, leverage temporal dependencies within clinical histories for more comprehensive recommendations. For clinical decision-making, Raza and Ding \cite{raza2024twostage} propose a two-stage recommendation framework to address the intricacies and time constraints of clinical workflows. Liu et al. \cite{liu2023llm} enhance clinical decision support systems by integrating Large Language Models (LLMs), facilitating the interpretation of complex medical data. For tag recommendation, Wang et al. \cite{wang2023tag} develop a tag recommendation system to classify and identify patients across extensive medical datasets. In doctor recommendation, the rising use of online health Q\&A platforms has motivated researchers to develop intelligent recommendation systems. Tian et al. \cite{tian2019drgan} introduce the DRGAN model, based on GAN networks, to address doctor recommendation challenges, while Zheng et al. \cite{zheng2022ddr} propose DDR, a conversation-based model designed to overcome cold-start issues due to limited historical data and incomplete patient descriptions.

Despite the broad application of recommendation systems in various medical domains, there remains a significant gap in research focusing on recommending appropriate medical examinations.

%\vspace{-0.3cm}
\subsection{Sequential Recommendation}
In recent years, substantial efforts have been made to improve sequential recommendations through deep learning techniques. Existing efforts can be broadly categorized into CNN-based, RNN-based, and Transformer-based methods \cite{yin2024dataset}\cite{zhou2025dual}.

\textbf{CNN-based methods} \cite{tuan20173d,yan2019cosrec, tang2018personalized}  are adept at capturing short-term user preferences from recent interactions. Tang et al. \cite{ tang2018personalized} embed recent interaction items into representations and apply CNNs to dynamically model these representation sequences. Yuan et al. \cite{yuan2019simple} introduce residual structures NextItNet architecture for handling longer interaction sequences. However, due to the limitation of localized characteristics, CNN-based methods fail to capture the long-term preferences of users.
\textbf{RNN-based methods}\cite{hidasi2015session,hidasi2018recurrent,qin2017dual,li2017neural} have emerged to address these limitations. Hidasi et al.\cite{hidasi2015session}introduce a Gated Recurrent Unit (GRU) into the sequential recommendation task to model the long-term interactions of users. Additionally, attention mechanisms have been incorporated into RNNs by Qin et al. \cite{qin2017dual} and Li et al. \cite{li2017neural}, further improving the modeling of long interaction sequences. 
Recently, \textbf{Transformer-based methods} \cite{hou2022core,zhou2020s3,kang2018self,sun2019bert4rec, xu2019graph, fan2021lighter} has received extensive research attention as their effectiveness in sequential modeling. For instance, Kang et al. \cite{kang2018self} are the first to introduce a Transformer for sequential recommendation tasks. Building on this, Sun et al. \cite{sun2019bert4rec} propose a bidirectional Transformer to account for the bidirectional influence of historical interactions. Moreover, Xu et al. \cite{xu2019graph} combine Graph Neural Networks (GNNs)\cite{wang2023diffusion} with self-attention mechanisms to enhance global sequential modeling. Recently, Fan et al. \cite{fan2021lighter} propose low-rank decomposition to reduce the computational complexity of self-attention mechanisms.

Compared to traditional user-item sequential recommendation, medical examination recommendation involves more complex interactions between patients and examinations. This makes existing sequential recommendation methods difficult to directly apply to medical examination recommendations for precise recommendations.

\subsection{Medical Domain Dataset}
With the widespread adoption of information technology in the medical field, electronic medical records (EMRs) have been widely applied in major medical institutions, providing a rich data foundation for related research, such as diagnosis prediction and drug recommendation tasks.

Diagnosis prediction is a critical task in the field of smart healthcare, and today there are numerous foundational datasets supporting its rapid development. For example, the UCI\footnote{https://archive.ics.uci.edu/datasets} Machine Learning Repository, proposed by the University of California, Irvine, contains a large number of medical-themed datasets, including those related to Heart Disease, Breast Cancer, Thyroid Disease, and others. Similarly, the KEEL (Knowledge Extraction based on Evolutionary Learning)\footnote{https://sci2s.ugr.es/keel/datasets.php} database also provides a significant number of datasets for medical classification and diagnosis, including Appendicitis, Hepatitis, Dermatology data, and more. Furthermore, the claims data of Health Insurance Review and Assessment Service (HIRA) \cite{kim2014guide} has recorded the diagnoses, treatments, procedures, surgical histories, and prescription drug information of nearly 50 million patients, offering a valuable resource for healthcare service research. The Medicaid\footnote{https://www.medicaid.gov/} dataset comprises insurance claims from the years 2011 and 2012, encompassing 99,159 patients and 2,034,485 visits.

In parallel, drug recommendation, a crucial medical task enhancing prescription precision relies on rich datasets supporting research endeavors. Large-scale clinical repositories like MIMIC-III \cite{johnson2016mimic}, MIMIC-IV \cite{johnson2023mimic}, and the eICU Collaborative Research Database \cite{pollard2018eicu}, curated by PhysioNet \cite{goldberger2000physiobank}, offer detailed patient medication profiles essential for drug recommendation studies. These repositories not only include medication data but also encompass demographic details, laboratory findings, disease diagnoses, and other clinical insights, facilitating a broad spectrum of medical analyses, including diagnosis prediction \cite{gupta2018using,xu2023seqcare}.

Many existing medical datasets cover diverse clinical domains but omit fully structured examination records, which limits their applicability to examination recommendations. To address this gap, we develop MeExam, a temporally organized, heterogeneous dataset specifically tailored for medical examination recommendations.

% Despite the availability of diverse medical datasets across various medical domains, the lack of complete structured examination data within them hinders their direct transferability to examination recommendation tasks. To address this gap, we propose the development of a temporal heterogeneous medical examination recommendation dataset.

\section{Preliminaries}
In this section, we first formally define the medical examination recommendation task and then introduce the patient EHR heterogeneous graph, which serves as the foundation for our proposed method. These definitions establish the necessary mathematical notation and problem formulation for the subsequent technical discussions.

\textbf{Definition 1: Medical examination recommendation}. Medical examination recommendation aims to use patients' sequential data and personal attribute data as inputs to assess their health risk status and recommend the most appropriate examination items for the next phase as shown in Figure~\ref{fig:1 introduction}. Suppose $U = \{u_1, \dots, u_{|U|} \}$ represents a set of patients. The corresponding sequential data is denoted as $S = \{s_1, \dots, s_{|U|}\}$. For each patient, their sequential data $s_u$ consists of various heterogeneous entities $s_u = \{o_1, \dots, o_t\}$, where $o_j \in O$ and $O$ represents a set of medical entities including diseases $D$, symptoms $P$, and examinations $C$. Additionally, the personal attribute data include age and gender information, which are represented as $B = \{b_1, \dots, b_{|U|}\}$  and $G = \{g_1, \dots, g_{|U|}\}$, respectively. Given the above context and notations, the medical examination recommendation problem can be defined as follows:

\begin{itemize}
    \item \textbf{Input:} Patient sequential data $s_u = \{o_1, \dots, o_p\}$ and the corresponding personal attribute information $b_u$ and $g_u$.
    \item \textbf{Output:} Predict the examination $o_{p+1}$ $(o_{p+1} \in C)$ that patient $u$ should engage with at the next phase $p+1$.
\end{itemize}

\begin{figure*}[t]
    \centering
    \includegraphics[width=0.9\linewidth]{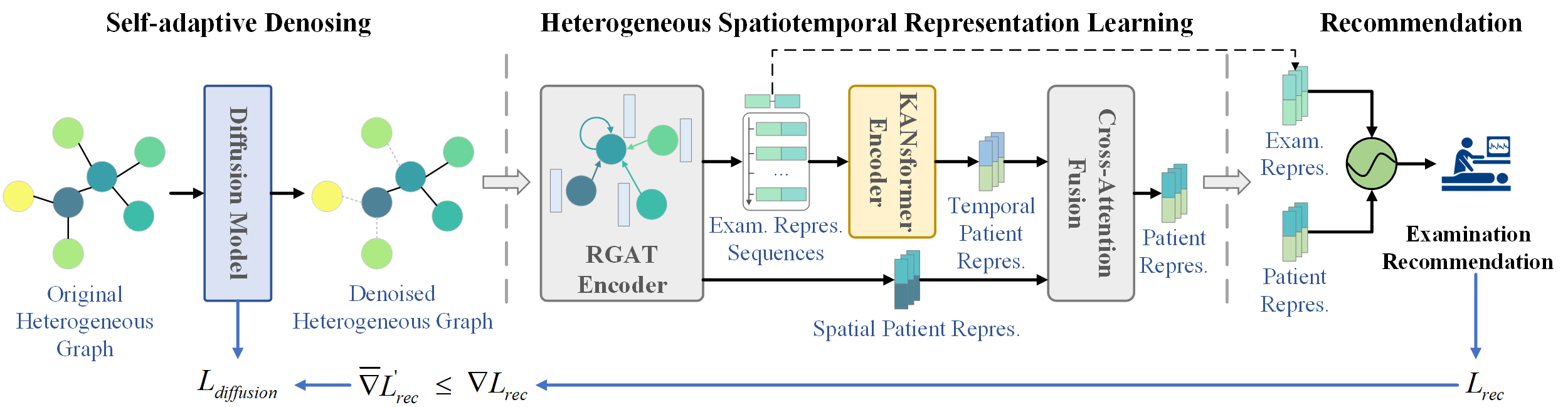}
    \caption{The overview of proposed Diffusion-driven SpatioTemporal Graph KANsformers (DST-GKAN) model. It comprises two stages. (1) Task-adaptive Denoising: the task-adaptive diffusion model is employed for denoising. (2) Heterogeneous Spatiotemporal Representation Learning:  Relation-aware Graph Attention Network (RGAT), KANsformer and Cross-Attention fusion are performed for feature representation.}
    \label{fig:2 framework}
\end{figure*}

% \subsection{Definition 2: Patient EHR Heterogeneous Graphs}
\textbf{Definition 2: Patient EHR heterogeneous graph}. Considering the heterogeneity of examination recommendation scenario of patients $U$ and medical entities $O$, we construct a binary EHR Heterogeneous Graph (HG) denoted as $\mathcal{G}$ to represent the collaborative signals between patients and medical entities.
We denote the EHR HG as $\mathcal{G} = (\mathcal{V}, \mathcal{E})$, consisting of the node set $\mathcal{V}$ and the edge set $\mathcal{E}$. The node set $\mathcal{V} =U \cup O \cup B \cup G$ includes six types of nodes: patients, diseases, symptoms, examinations, age, and gender. Formally, the HG $\mathcal{G}$ can be represented as an adjacency matrix $A_{\left| U \right| \times (\left| \mathcal{V} \right| - \left| U \right|)} = \{ a_{ui} | u \in \mathcal{V}, i \in \mathcal{V} \}$. Here, $a_{ui} \in A$ is equal to 1 if patient $u$ has interacted with examination item $i$, and 0 otherwise. $a_{ui} = 1$ means that there is a connecting edge between $u$ and $i$ in the HG $\mathcal{G}$.

\section{The Proposed Method}
% \subsection{Overview}
In this section, we present the technical design of the DST-GKAN model, accompanied by the overall model architecture depicted in Fig~\ref{fig:2 framework}. The structure of this model consists of three main components: task-adaptive denoising with diffusion model, heterogeneous spatiotemporal representation learning, and medical examination recommendation.

\subsection{Task-Adaptive Denoising with Diffusion Model}

% Considering additional knowledge in the HG can enhance the model's capacity to evaluate patients' health status. However, 
Considering that not all information in HG is pertinent to the downstream recommendation tasks, other knowledge relationships (task-irrelevant relationships) can introduce noise that may mislead the model’s inference of patients' health status. Inspired by the effectiveness of diffusion models \cite{jiang2024diffkg, wang2023diffusion} in data generation from noisy inputs, we propose a task-adaptive denoising approach with diffusion model to generate a recommendation-relevant subgraph $\mathcal{G}_d$ from the original HG $\mathcal{G}$, thereby mitigating the impact of task-irrelevant information. The approach consists of two components: diffusion-based denoising and task-adaptive mechanism.

\begin{figure}[ht]
    \centering
    \includegraphics[width=1\linewidth]{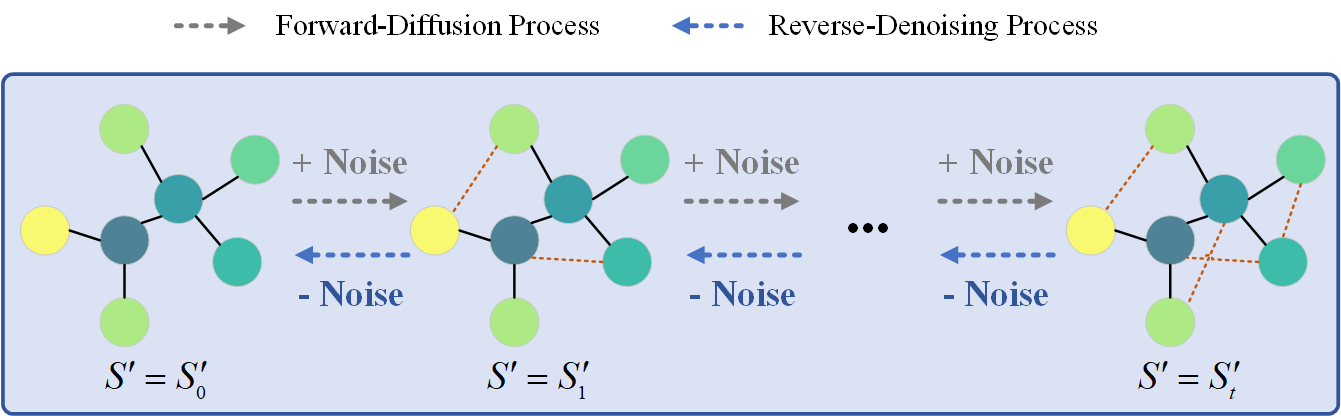}
    \vspace{-0.3cm}
    \caption{ The illustration of forward (noising) and reverse (denoising) processes in our diffusion models. }
    \label{fig:3 diffusion}
    % \vspace{0.3cm}
\end{figure}

% To achieve this, the model is trained to identify true relationships between items and entities in a knowledge graph that has been corrupted by a noise diffusion process. Our method employs a forward process that gradually introduces noise to the relations in the knowledge graph, simulating the corruption of relations. Then, through iterative learning, we aim to recover the original relations in the knowledge graph. This iterative denoising training enables DiffKG to model complex relation generation procedures and reduce the impact of noisy relations. Ultimately, the restored relation probabilities are utilized to reconstruct the subgraph from the original knowledge graph $G_k$.

\subsubsection{Diffusion-Based Denoising}

The diffusion-based denoising consists of two essential processes: the forward process and the reverse process, as illustrated in Fig~\ref{fig:3 diffusion}. The forward process incrementally injects noise into the original data samples, progressively transforming them into Gaussian noise. In contrast, the reverse process recovers the original samples step-by-step by predicting and removing the added noise.

\textbf{Forward Process}. The original structure of the original HG $\mathcal{G}$ is corrupted by adding Gaussian noises step by step. Specifically, we initialize the initial state $x_0$ as the original adjacency matrix $A$, i.e., $x_0 = A$. The forward process then constructs $x_{0:T}$ in a Markov chain by gradually adding Gaussian noise in $T$ steps. We parameterize the transition from $x_{t-1}$ to $x_t$ as:

\begin{equation}
q(x_t \mid x_{t-1}) = \mathcal{N}(x_t; \sqrt{1-\beta_t} x_{t-1}, \beta_t I),
\label{eq:forward1}
\end{equation}
where $t \in 1, \cdots, T$ represents the diffusion step, $\mathcal{N}$ denotes the Gaussian distribution, and $\beta_t \in (0,1)$ controls the scale of the Gaussian noise added at each step $t$. As $T \to \infty$, the state $x_T$ converges towards a standard Gaussian distribution. By utilizing the reparameterization trick and taking the advantage of the additivity property of two independent Gaussian noises, we can directly derive the state $x_T$ from the initial state $x_0$. Formally, we describe this process as follows:

\begin{equation}
q(x_t \mid x_0) = \mathcal{N}(x_t; \sqrt{\bar{\alpha}_t} x_0, (1 - \bar{\alpha}_t) I), \bar{\alpha}_t = \prod_{t'=1}^t (1 - \beta_{t'}).
\label{equ2}
\end{equation}
Based on the Eq. \ref{equ2}, $x_t$ can be reparameterized as follows:
\begin{equation}
x_t = \sqrt{\bar{\alpha}_t} x_0 + \sqrt{1 - \bar{\alpha}_t} \epsilon, \quad \epsilon \sim \mathcal{N}(0, I).
\end{equation}
To regulate the addition of noise in $x_{1:T}$, we incorporate a linear noise scheduler that implements $1 - \bar{\alpha}_t$ using three hyperparameters: $s$, $\alpha_{\text{low}}$, and $\alpha_{\text{up}}$. The linear noise scheduler is defined as follows:

\begin{equation}
1 - \bar{\alpha}_t = s \cdot [\alpha_{\text{low}} + \frac{t-1}{T-1} (\alpha_{\text{up}} - \alpha_{\text{low}})], \quad t \in \{1, \cdots, T\},
\end{equation}
where $s \in [0, 1]$ controls the noise scales, while $\alpha_{low} < \alpha_{up} \in (0, 1)$ sets the upper and lower bounds for the added noises. Besides, $T$ denotes the maximum time step.  

\textbf{Reverse Process}. The diffusion model learns to remove the added noise from $x_t$ via neural networks in order to recover $x_{t-1}$. Starting from $x_t$, the reverse process gradually reconstructs the relations within the HG through the denoising transition step. The denoising transition step is outlined as follows:

\begin{equation}
    p_\theta(x_{t-1} \mid x_t) = \mathcal{N}(x_{t-1}; \mu_\theta(x_t, t), \Sigma_\theta(x_t, t)).
    \label{eq:subgraph}
\end{equation}
Here, we utilize neural networks parameterized by $\theta$ to generate the mean $\mu_\theta(x_t, t)$ and covariance $\Sigma_\theta(x_t, t)$ of a Gaussian distribution.

\textbf{Optimization.} Similar to VAEs, DMs are optimized by maximizing the Evidence Lower Bound (ELBO) of the likelihood of observed input data $x_0$. We aim to maximize the ELBO of observed user interaction $\mathcal{H}$ (i.e. HG) to learn $\theta$ in the reverse process:

% To learn $\theta$, DiffRec aims to maximize the ELBO of observed user interaction $\mathbf{x_0}$:

\begin{equation}
\begin{aligned}
    \log p(x_0) \geq {} & \mathbb{E}_{q(x_1|x_0)} [ \log p_\theta(x_0|x_1) ] \\
    & - \sum_{t=2}^{T} \mathbb{E}_{q(x_t|x_0)} \left[ \text{D}_{\text{KL}} (q(x_{t-1}|x_t, x_0) || p_\theta(x_{t-1}|x_t)) \right].
\end{aligned}
\label{eq:logp}
\end{equation}
The first term (reconstruction term) measures the recovery probability of $x_0$, representing the ability of the model to reconstruct the original HG. The second term (denoising matching term) regulates the recovery of $x_{t-1}$ for $t$ ranging from 2 to $T$ in the reverse process.

For the first term, the $\mathcal{L}_1$ is defined as the negative of the reconstruction term in Eq.~(\ref{eq:logp}) and it can be calculated as follows:
\begin{equation}
\begin{aligned}
   \mathcal{L}_1 
    &\triangleq -\mathbb{E}_{q(x_1|x_0)} \left[ \log p_\theta(x_0|x_1) \right] \\
    &= \mathbb{E}_{q(x_1|x_0)} \left[ \left\| \hat{x}_\theta(x_1, 1) - x_0 \right\|^2_2 \right]
\end{aligned}
\end{equation}
where we estimate the Gaussian log-likelihood $\log p_\theta(x_0|x_1)$ through unweighted $-\| \hat{x}_0(x_1, 1) - x_0 \|^2_2$ as discussed in [22].

The second denoise matching term $\mathcal{L}_t$ is detailed as follows:
\begin{equation}
    \mathcal{L}_t = \mathbb{E}_{q(x_t|x_0)} \left[ \frac{1}{2} \left( \frac{\bar{\alpha}_{t-1}}{1 - \bar{\alpha}_{t-1}} - \frac{\bar{\alpha}_t}{1 - \bar{\alpha}_t} \right) \| \hat{x}_\theta(x_t, t) - x_0 \|^2_2 \right],
\label{equ8}
\end{equation}
where $\hat{x}_\theta(x_t, t)$ is the predicted $x_0$ based on $x_t$ and $t$. To calculate Eq.~(8), we implement $\hat{x}_\theta(x_t, t)$ by neural networks. Specifically, we instantiate $\hat{x}_\theta(\cdot)$ via a Multi-Layer Perceptron (MLP) that takes $x_t$ and the step embedding of $S'_t$ as inputs to predict $x_0$. According to Eq.~\ref{equ8}, ELBO in Eq.~\ref{eq:logp} can be formulated as $-\mathcal{L}_1 - \sum_{t=2}^{T} \mathcal{L}_t$. Hence, to maximize the ELBO, we can optimize $\theta$ in $\hat{x}_\theta(x_t, t)$ by minimizing $\sum_{t=1}^{T} \mathcal{L}_t$. Specifically, we sample step $t$ under a uniform distribution of $t \sim \mathcal{U}(1, T)$ to optimize $\mathcal{L}(x_0, \theta)$. Formally, the ELBO loss $\mathcal{L}_{\text{ELBO}}$ is shown below:

\begin{equation}
    \mathcal{L}_{\text{ELBO}} = \mathcal{L}(x_0, \theta) = \mathbb{E}_{t \sim \mathcal{U}(1, T)} \mathcal{L}_t
    \label{elbo}
\end{equation}

\subsubsection{Task-Adaptive Mechanism}
To achieve task-driven denoising, we introduce a task-adaptive function $r(\cdot)$ that guides the diffusion paradigm to generate task-relevant subgraphs. It is defined as:

\begin{equation}
r(\mathcal{L} _{rec}) = 
\begin{cases} 
1 & \text{if } \nabla \mathcal{L}_{\text{rec}} > \overline{\nabla }\mathcal{L}_{\text{rec}}', \\
\epsilon & \text{otherwise}
,
\end{cases}
\end{equation}

\noindent where $\mathcal{L}_{\text{rec}}$ represents the loss function of the downstream recommendation tasks. $\nabla \mathcal{L}_{\text{rec}}$ denotes the difference between the current training step and the last step, while $\overline{\nabla } \mathcal{L}_{\text{rec}}'$ represents the averaged change of $\mathcal{L}_{\text{rec}}$ over the previous $\delta > 1$ steps. If $\nabla {\mathcal{L}_{rec}} \le \bar \nabla {\mathcal{L}'_{rec}}$, it indicates that the generated heterogeneous subgraph in the current epoch is beneficial for downstream task training, leading to a faster reduction in the loss function. In this case, we suppress the training of the diffusion model with a threshold $\varepsilon < 1$ to preserve the current graph structure. Conversely, if $\nabla {\mathcal{L}_{rec}} > \bar \nabla {\mathcal{L}'_{rec}}$, we resume training to guide the diffusion model in reconstructing a subgraph more suitable for the downstream recommendation task. The final task-adapted loss function can be defined as:
\begin{equation}
    {\mathcal{L'}_{ELBO}} = r\left( {{\mathcal{L}_{rec}}} \right){\mathcal{L}_{ELBO}},
    \label{elob2}
\end{equation}

\noindent where $\mathcal{L}_{ELBO}$  represents the original loss function of the diffusion model. This learning mechanism dynamically adjusts the training of the diffusion model based on the impact of the currently generated heterogeneous subgraph on the downstream task, thereby facilitating an adaptive and task-specific training process.

\subsection{Heterogeneous Spatiotemporal Representation Learning}
The heterogeneous spatiotemporal representation learning consists of three key phases: 1) A Relation-aware Graph Attention Network (RGAT) is employed to aggregate spatial features among nodes, capturing the relational dependencies within the denoised subgraph; 2) A KANsformer is devised as a temporal encoder to extract the temporal dynamic features of patients over time; 3) Feature fusion is accomplished through a cross-attention mechanism, which integrates the learned spatial and temporal features to obtain the comprehensive representations of patients for precise examination recommendations.

% The Heterogeneous Spatiotemporal Representation Learning is composed of three phases: 1) Relation-aware Graph Attention network (RGAT) aggregates the features of spatial nodes; 2) KANsformer as sequence encoder to obtain the temporal characteristics of patients; 3) Feature Fusion through Cross Attention: This process combines the learned spatial and temporal features of patients to forecast the examination items.

\subsubsection{RGAT as Spatial Encoder}
To fully leverage the complex heterogeneous relationships within the heterogeneous graph, we implement a heterogeneous relation-aware encoding module inspired by the graph attention mechanism \cite{jiang2024diffkg, yang2023debiased}. This module effectively captures the diverse relationships inherent in the global connecting structure of the collaborative HG. During the information propagation process, let $N_m$ denote the set of all neighboring entities connected to entity $m$ in the denoised heterogeneous subgraphs $\mathcal{G}_d$ through relation $r_{m,n}$. The embeddings for entities $m$ and $n$ are represented as $x_m \in \mathbb{R}^d$ and $x_n \in \mathbb{R}^d$, respectively, which are initialized as learnable parameters. We estimate the attention relevance specific to the entities and relations, denoted as $\alpha(m, r_{m,n}, n)$, to capture the unique semantic relationships between entities $m$ and $n$. The specific information propagation mechanism can be described as follows:
\begin{equation}
    x_m^l = Drop(Norm(x_m^{l-1} + \sum_{i \in N_m} \alpha(m, r_{m,n}, n)^{l-1} x_n^{l-1}))
    \label{eq:rgat1}
\end{equation}
\begin{equation}
    \alpha(m, r_{m,n}, n)^{l-1} = \frac{\exp(\sigma(r_{m,n}^T W [x_m^{l-1} \parallel x_n^{l-1}]))}{\sum_{i \in N_m} \exp(\sigma(r_{m,n}^T W [x_m^{l-1} \parallel x_n^{l-1}]))}
    \label{eq:rgat2}
\end{equation}
where $\parallel$ denotes the concatenation operation, $W$ denotes a learnable weight, and $\sigma$ denotes $LeakyReLU\left( \cdot \right)$. Here, to avoid overfitting, we employ dropout functions ($Drop\left( \cdot \right) $) and normalization ($Norm\left( \cdot \right)$) for the node representation learning at each layer, following the processing approach outlined in \cite{yang2023debiased}. Ultimately, we fuse the entity representations from all layers to form the global collaborative spatial representations, which can be represented as $e_{m}^{spa} = \sum_{l=1}^{L} x_m^l$. Here, $e_{m}^{spa}$ denotes the representations of any entity $m$ on the subgraphs $\mathcal{G}_d$, including patient representation $e_{u}^{spa}$ and examination representation $e_{i}^{spa}$. And $L$ denotes the number of RGAT layers.

\subsubsection{KANsformer as Temporal Encoder}
\begin{figure}[ht]
    \centering
    \includegraphics[width=0.8\linewidth]{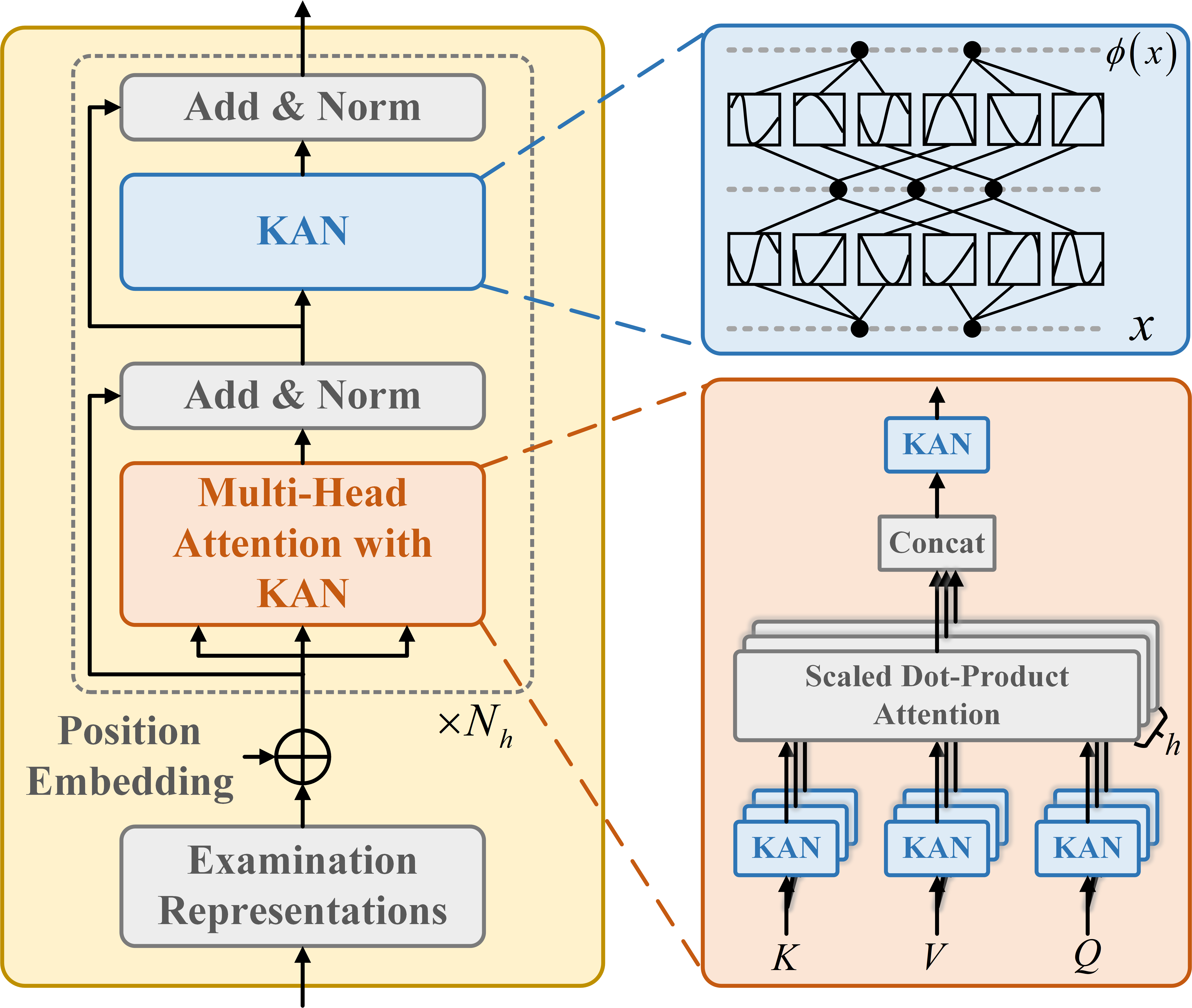}
    % \vspace{-0.3cm}
    \caption{The framework of the proposed KANsformer. The KANs is integrated into the Transformer framework.}
    \label{fig:4 KANsformer}
    % \vspace{0.3cm}
\end{figure}

Building upon the global collaborative spatial features, it is essential to extract the temporal dependencies of patients' medical history for health status diagnosis. While Transformers \cite{ashish2017attention} excel at capturing temporal patterns through self-attention mechanisms, their conventional feed-forward layers face limitations in efficiently learning complex functional mappings. Kolmogorov–Arnold Networks (KANs) \cite{liu2024kan} shown significant potential to address this by replacing fixed activation functions with learnable spline-based structures, enabling adaptive nonlinear feature learning. Therefore, we propose a novel architecture, termed KANsformer, which integrates KANs into the Transformer framework. As shown in Fig~\ref{fig:4 KANsformer}, KANsformer retains the self-attention mechanism of Transformer to model global temporal dependencies but replaces its feed-forward layers with KAN modules. This integration enhances the network's ability to approximate intricate temporal dynamics in sequential patient data while maintaining computational efficiency.

Firstly, we incorporate position embedding $b_i$ into the global collaborative spatial representations $e_i^{spa}$ of item $i$, thereby allowing the patient's sequential data to be expressed as $h_i = e_i^{spa} \oplus b_i$. And we can obtain the item representation sequence $H_u^0 = \{h_1, h_2, ..., h_i , ..., h_p\}$ of each patient $u$ as the input of our proposed KANsformer. The KANsformer retains the core component of the Transformer architecture, namely the multi-head self-attention mechanism, which can be formulated as follows:
% \begin{equation}
%     H_u^l = \mathbin\Vert_{j=1}^{N_h} \phi_V^j (\hat{\alpha}^j H_u^l);
%     \hat{\alpha}^j = \frac{\phi_Q^j (H_u^l) \phi_K^j (H_u^l)}{\sqrt{d/H}},
%     \label{eq:kans1}
% \end{equation}
\begin{equation}
    H_u^l = \mathop \parallel \limits_{j = 1}^{{N_h}} \phi _V^j({\hat \alpha ^j}H_u^l);
    \hat{\alpha}^j = \frac{\phi_Q^j (H_u^l) \phi_K^j (H_u^l)}{\sqrt{d/H}},
    \label{eq:kans1}
\end{equation}

\begin{equation}
    H_u^{l+1} = \text{GELU}(\phi_2 (\phi_1 (H_u^l))),
    % h_{u_T} = H_{u_{|t|}}^{L_2},
    \label{eq:kans2}
\end{equation}
where $l$ and $N_h$ represents the number of multi-head self-attention blocks and the number of heads, respectively. The attention value for the $j$-th head is denoted as $\hat{\alpha}^j$. The single-layer efficient KANs function is indicated as $\phi(\cdot)$, where $\phi_Q^j$, $\phi_K^j$, and $\phi_V^j$ correspond to the $j$ efficient KANs functions associated with the queries, keys, and values, respectively, facilitating the feature mapping transformation. Analogous to the Feed-forward Network (FFN) in the Transformer architecture, we employ a two-layer efficient KANs function to further refine the temporal features of patients. Following the approach outlined in \cite{sun2019bert4rec}, we utilize the final vector from the output sequence of the KANsformer as the patient's local temporal features $e_u^{tem}$ for subsequent calculations.

\subsubsection{Feature Fusion by Cross-Attention}
To fully integrate spatial and temporal features, we introduce the cross-attention mechanism to achieve deeper feature fusion.
Cross-attention has demonstrated promising results in feature fusion tasks in previous studies \cite{ashish2017attention, li2024multimodal, sun2023attention, shen2024icafusion, zheng2023casf, chen2021crossvit, 10597701}. Its primary purpose is to leverage the self-attention mechanism to derive contextual information from the Query (Q) features that enhance the Key-Value (K-V) features, thereby refining the final representations. In this work, we set the collaborative spatial features as Q and the temporal features as V-K, using cross-attention to embed the spatial context features into the temporal features. This enables spatiotemporal semantic alignment, enriching the personalized representation for each patient. 

First, we apply multiple linear projection layers to separately transform the patient's collaborative spatial features $e_u^{spa}$ and temporal features $e_u^{tem}$ into their corresponding query $Q$, key $K$, and value $V$ features. This step enables the effective mapping of both global and local features into a unified space, facilitating the subsequent feature alignment through cross-attention.
\begin{equation}
    Q_{u_s} = W^Q e_u^{spa}, \quad K_{u_T} = W^K e_u^{tem}, \quad V_{u_T} = W^V e_u^{tem},
    \label{eq:cross1}
\end{equation}
where $W^Q$, $W^K$, and $W^V$ denote the learnable projection matrices used to map the features into the query, key, and value spaces, respectively. And then, the cross-attention mechanism ($Cro\_Attn$) for feature fusion is computed as follows:
\begin{equation}
    e_u^{final} = Cro\_Attn(Q_{u_s}, K_{u_T}, V_{u_T}),
    \label{eq:cross2}
\end{equation}
\begin{equation}
    Cro\_Attn(Q, K, V) = softmax\left(\frac{Q K^T}{\sqrt{d_k}}\right) V,
    \label{eq:cross3}
\end{equation}
where $d_k$ represents the embedding dimension of the features in the query, key, and value spaces. And $e_u^{final}$ denotes the final learned spatiotemporal status representation of the patient.

\subsection{Recommendation and Model Training}
For the medical examination recommendation task, we use the widely adopted inner product~\cite{10597799, 10598015} to generate recommendation results. Formally, for a given candidate examination entity $i \in C$, the recommendation score can be computed as follows:
\begin{equation}
    \hat{y}_{ui} = e_u^{final} {(e_i^{spa})}^T,
    \label{eq:score}
\end{equation}
where vector $\hat{y}_u = (\hat{y}_{u1}, \hat{y}_{u2}, ..., \hat{y}_{u|C|})$ represents the score vector of patient $u$ for each candidate examination item. 

Subsequently, since the task of medical examination recommendation is essentially a sequential prediction problem, we employ the cross-entropy loss \cite{mao2023cross} as our objective loss function $\mathcal{L}_{\text{Rec}}$, which is formulated as follows:
% We then use $\mathcal{L}_{\text{Rec}}$ as cross-entropy loss \cite{mao2023cross} for the recommendation task. Since the task of medical examination recommendation is essentially a sequential prediction problem, we employ the cross-entropy loss \cite{mao2023cross}. The objective loss function is formulated as follows: 

%The training of our model consists of two primary components: training for the recommendation task and training for Task-adaptive Denoise. Therefore, the integrative optimization loss for the recommendation task is:

% Finally, the model parameters are optimized using cross-entropy loss,
% \begin{equation}
%     L_\text{Rec} = -\sum_U \sum_{c=1}^{|C|} \left( y_{ui} \log(\hat{y}_{ui}) + (1 - y_{ui}) \log(1 - \hat{y}_{ui}) \right) + \lambda \| \theta \|_2,
% \end{equation}
% here, $y_u$ denotes the one-hot encoded vector representing the ground truth examination item for patient $u$'s next interaction, $\theta$ represents all model parameters, and $\| \cdot \|_2$ refers to the $L_2$ norm. $\lambda$ is used to control the regularization strength and mitigate overfitting.

% \subsection{The Model Training}

%\begin{equation}
%    \mathcal{L}_{\text{fin}} = \mathcal{L}_{\text{Rec}} + %\mathcal{L}_{ELBO},
%\end{equation}
%where $\mathcal{L}_{\text{Rec}}$ is cross-entropy loss \cite{mao2023cross} for recommendation task. Since the task of medical examination recommendation is essentially a sequential prediction problem, we employ the cross-entropy loss \cite{mao2023cross}. The objective loss function is formulated as follows:
\begin{equation}
    \mathcal{L}_{\text{Rec}} = -\sum_{(u,i) \in Q_{\text{UC}}}  y_{ui} \log \hat{y}_{ui} + (1 - y_{ui}) \log (1 - \hat{y}_{ui})  + \lambda \|\theta\|_2,
    \label{eq:lrec}
\end{equation}
where $Q_{\text{UC}} = \{(u, i) \mid y_{ui} = 1, u \in U, i \in C \}$ denotes the training data. $\theta$ represents all the model parameters, and $\| \space \|_2$ denotes the L2 norm. $\lambda$ is utilized to control the regularization strength.

%For training step, as shown in Algorithm \ref{algorithm:training}, we divide the training process of DST-GKAN into two stages: (1) obtain the denoised subgraph related to the downstream recommendation task by training the task-adaptive diffusion model based on Eq.~(1) to Eq.~(3). (2) we calculate the recommendation score and loss function to update the parameters of the model based on Eq.~(13) and Eq.~(14). 

\begin{algorithm}
\caption{Training Procedure of DST-GKAN.}
\label{algorithm:training}
\begin{algorithmic}[1]  % 启用行号并从1开始
    \State \textbf{Input:} Training data set $Q_{\text{UC}}$ with the personal information of patients, training epochs $N$, and learning rate $l$;
    \State \textbf{Initialize:} Parameters $\theta$;
    \State Construct the interaction sequences $S$ of patients according to $Q_{\text{UC}}$;
    \State Construct the global collaborative heterogeneous graph $\mathcal{G}$ according to all heterogeneous entities in sequences $S$ with the personal information;
    \State \textbf{Training Stage 1}
    \For{$\text{epoch} = 1$ to $N$}
    %\State \textbf{for} $epoch = 1$ to $N$ \textbf{do}
        \State Input $\mathcal{G}$ to the task-adaptive diffusion model;
        \State Output denoised subgraph $\mathcal{G}_d$ based on Eq.~(\ref{eq:subgraph});
        \State Optimize the loss $\mathcal{L}'_{\text{ELOB}}$ based on Eq.~(\ref{elob2})
    \EndFor
    \State \textbf{Training Stage 2}
    \For{$\text{epoch} = 1$ to $N$}
        \State Input $\mathcal{G}_d$ to the RGAT to obtain the global spatial features $h_{u_s}$ and $h_{i_s}$ of patients and items, respectively, based on Eq.~(\ref{eq:rgat1}) to Eq.~(\ref{eq:rgat2});
        \State Update $S$ according to $\mathcal{G}_d$;
        \State Input updated $S$ and $h_{i_s}$ to the KANsformer to obtain local temporal features $h_{u_t}$ of patients based on Eq.~(\ref{eq:kans1}) to Eq.~(\ref{eq:kans2});
        \State Input global spatial features $h_{u_s}$ and local temporal features $h_{u_t}$ to the cross-attention mechanism to obtain the final spatiotemporal features $h_u$ of patients based on Eq.~(\ref{eq:cross1}) to Eq.~(\ref{eq:cross3});
        \State Output the recommendation score $\hat{y}_{ui}$ based on Eq.~(\ref{eq:score});
        \State Optimize the loss $\mathcal{L}_{\text{Rec}}$ based on Eq.~(\ref{eq:lrec}) and feed back to the \textbf{Training Stage 1};
    \EndFor
\end{algorithmic}
\end{algorithm}

For the training step, as shown in Algorithm \ref{algorithm:training}, we divide the training process of DST-GKAN into two stages: 1) We obtain a denoised subgraph with the ELBO, reducing irrelevant noise in heterogeneous medical data. This stage uses a task-adaptive diffusion model as shown in Eq.~(\ref{eq:forward1}) to Eq.~(\ref{elob2}), to create subgraphs suited for downstream recommendation tasks. 2) We improve recommendation accuracy by optimizing the recommendation loss $\mathcal{L}_{\text{Rec}}$. Using Eq.~(\ref{eq:rgat1}) and Eq.~(\ref{eq:lrec}), we calculate the recommendation score and loss to update model parameters. This stage’s results are fed back to the first stage, producing denoised subgraphs more relevant to recommendations.

\subsection{The MeExam Dataset}
\subsubsection{Data Overview}
Despite the promising potential of medical examination recommendation, the field lacks publicly available benchmark datasets. To bridge this gap, we construct MeExam, a dataset designed to reflect real-world clinical workflows with both heterogeneous and temporal properties. The heterogeneous property arises from two main components: patient interactions and demographic information. Patient interactions comprise three types of entities, namely diseases, symptoms, and examinations, while demographic information includes age and gender. The temporal property is captured by organizing the entities in each patient's interactions in chronological order, based on anonymous medical notes.

\section{Data collection and preprocessing}
\begin{figure}[h]
    \centering
    \includegraphics[width=0.5\textwidth]{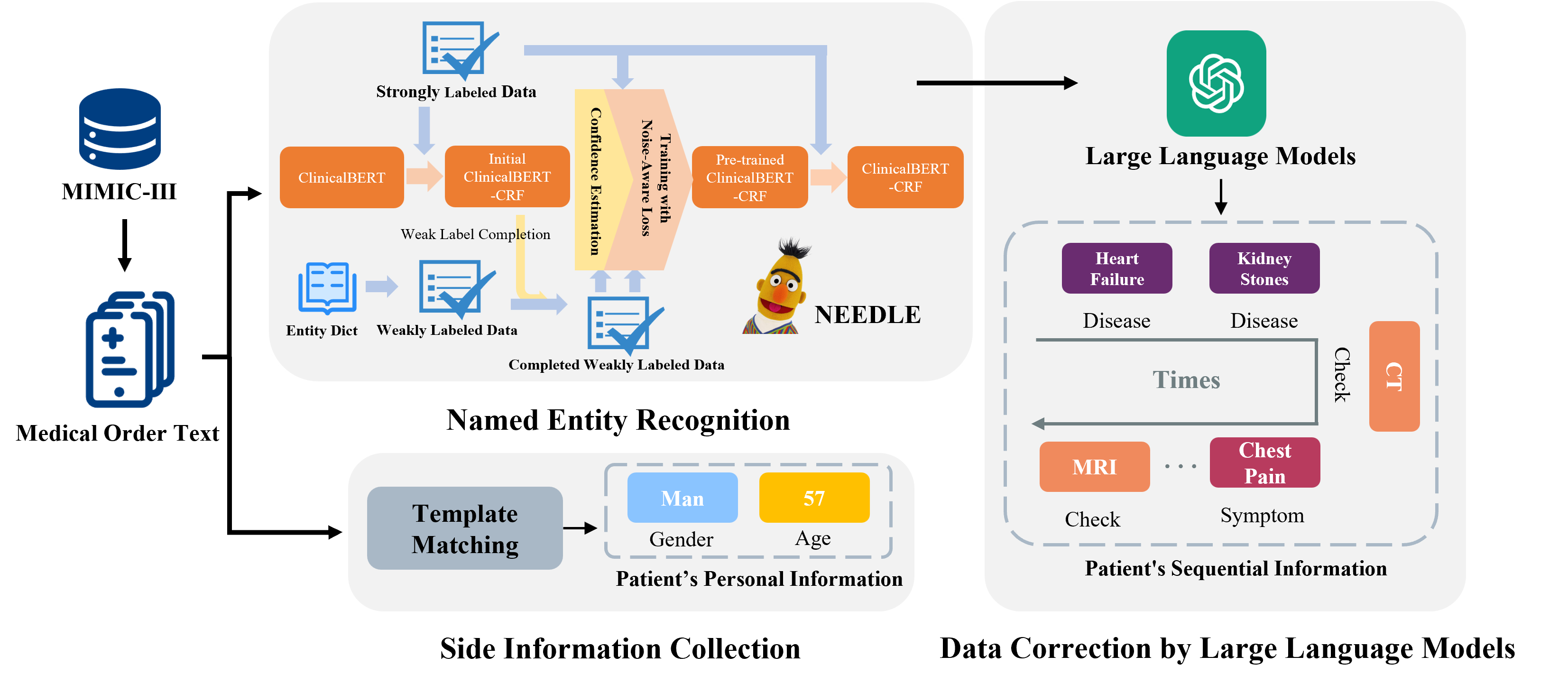}
    \caption{The framework of MeExam: i) The few-shot NER framework NEEDLE is used to extract the medical heterogeneous entities required for patient sequences from anonymous medical notes text data. ii) Side information, which consists of the patient's age and gender, is collected through template matching or retrieval methods. iii) Employ large language models to cleanse, disambiguate, and standardize the identified heterogeneous entities from NEEDLE.}
    \label{fig:dataset_model}
\end{figure}%

%\subsubsection{Data Construction Framework}
\subsection{Data Construction Framework}
As shown in Fig~\ref{fig:dataset_model}, the construction of the MeExam dataset from unstructured database files involves three stages: Named Entity Recognition, Side Information Collection, and Data Refinement with Large Language Models (LLMs). First, we apply named entity recognition techniques to extract key entities, including past medical history, current symptoms, and prior examinations, from anonymous medical notes in the MIMIC-III database. These entities are organized sequentially to preserve their temporal relationships. Second, recognizing the importance of demographic context in clinical decision-making, we enrich the dataset with side information, including each patient's age and gender, to support personalized modeling. Finally, we unleash the potential of LLMs to denoise and disambiguate the extracted entities, further enhancing data quality.
\begin{figure}[t]
  \centering
  \includegraphics[width=0.95\linewidth]{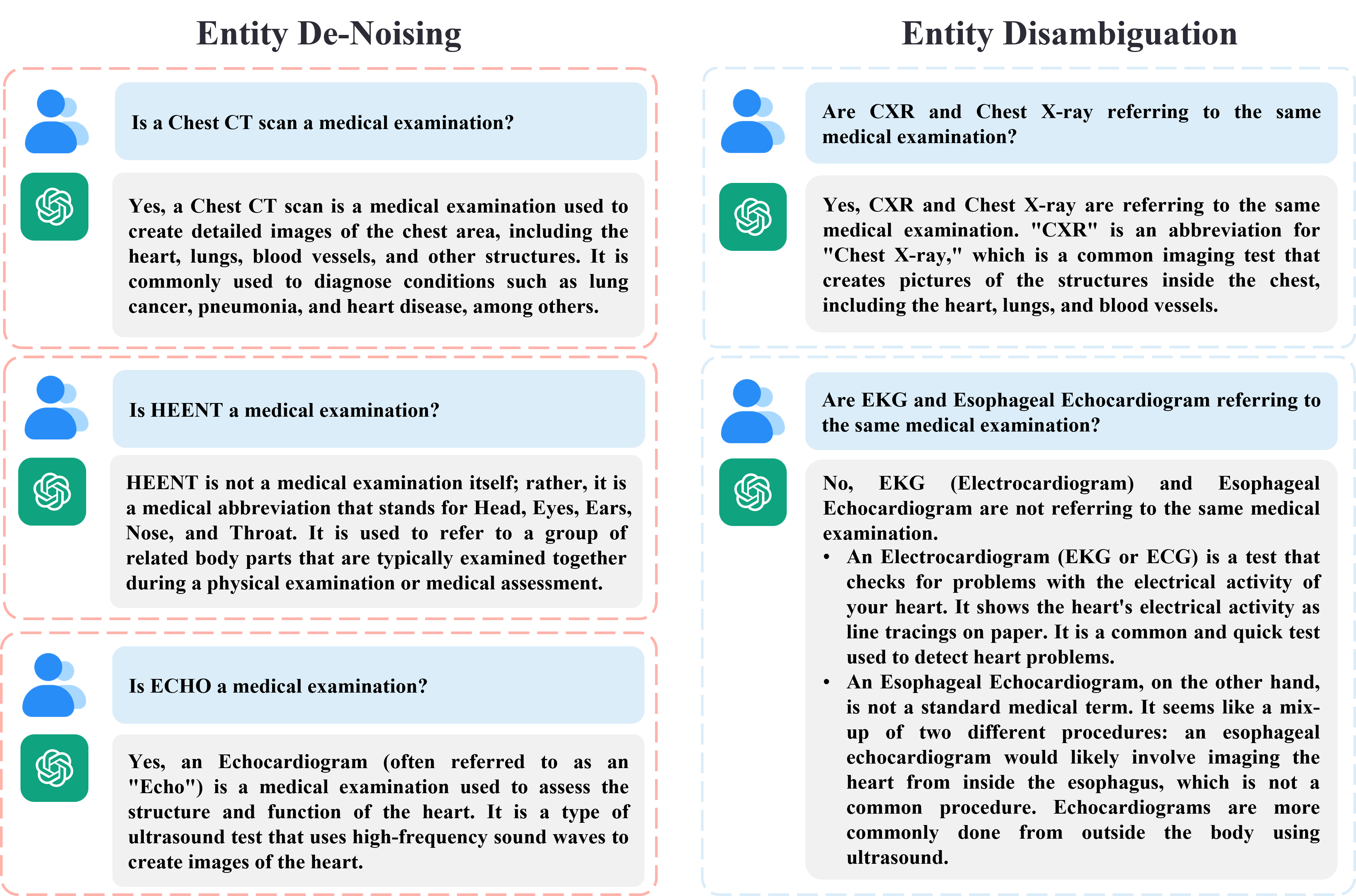}
  \caption{Example of prompt for LLMs. The blue background section represents the prompt we designed, while the gray background section shows the response from the LLMs.}
  \label{fig:prompt}
\end{figure}

\textbf{Named Entity Recognition.} The MIMIC-III database provides anonymous medical notes that chronologically document key clinical events, such as medical histories, symptom presentations, and examinations. However, these notes exist in an unstructured textual format. To extract structured data, we adopt a named entity recognition (NER) approach to automatically identify the heterogeneous entities needed for dataset construction. To minimize manual effort while maintaining high data quality, we leverage a few-shot NER framework NEEDLE \cite{jiang2021named}. NEEDLE follows a three-stage training process: it first generates a large set of weakly labeled data through dictionary matching, then refines these labels using confidence estimation, and finally mitigates the impact of label noise via a noise-aware loss function. To implement NEEDLE, we manually annotate approximately 5,000 entity labels covering 170,000 words and build three dictionaries for diseases, symptoms, and examinations via web crawling techniques. These dictionaries are used to generate a large volume of weakly labeled data, which are subsequently enhanced through NEEDLE's training procedure. Based on the enhanced labels, we select ClinicalBERT \cite{alsentzer-etal-2019-publicly} as the NER backbone model, given its pre-training on corpus in the medical field and strong ability to capture clinical context. The fine-tuned ClinicalBERT consistently delivers robust performance and generalization when applied to downstream tasks on the MIMIC-III dataset. In summary, we combine manually annotated strong labels and automatically generated weak labels to train the NER model. Using the NEEDLE framework and ClinicalBERT model, we successfully extract heterogeneous entities from the anonymous medical notes of 37,774 patients, organizing them into patient sequences based on the chronological order of clinical interactions.

\textbf{Side Information Collection.} In our constructed MeExam, side information includes each patient's gender and age. Gender is directly obtained by querying the patient registration table in MIMIC-III, which records gender, birth, and death dates linked to patient IDs. Although age is not explicitly structured, it frequently appears within the anonymous medical notes. To extract this information, we design a template-matching strategy based on common age expression patterns in the text. By analyzing the formats used in medical order documents, we create nearly ten templates, enabling the accurate extraction of age information for 37,774 patients.

\textbf{Data Refinement with LLMs.} Although NER extracts heterogeneous entities, the results still contain noise, such as incorrect entities or entity ambiguities. To cost-effectively improve data quality, we introduce LLMs, leveraging their extensive world knowledge and excellent semantic understanding capability \cite{achiam2023gpt,glm2024chatglm}. Through meticulously tailored prompts, we guide LLMs to perform entity denoising, disambiguation, and standardization. This process, combined with minimal human intervention, enables efficient and accurate refinement of the dataset. Fig~\ref{fig:prompt} shows examples of the designed prompts, demonstrating how LLMs effectively understand and perform instructions to enhance entity quality.

\begin{figure}[t]
  \centering
  \includegraphics[width=0.9\linewidth]{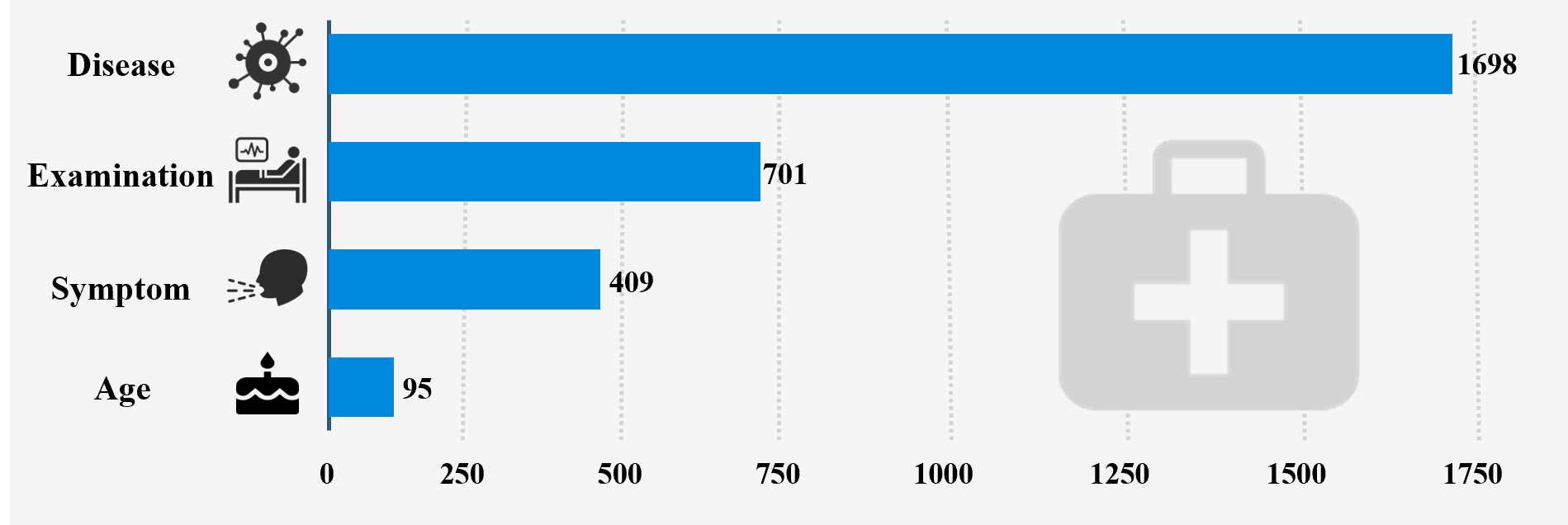}
  \caption{Entity type distribution.}
  \label{fig:Entity type}
\end{figure}

\begin{figure}[t]
  \centering
  \includegraphics[width=\linewidth]{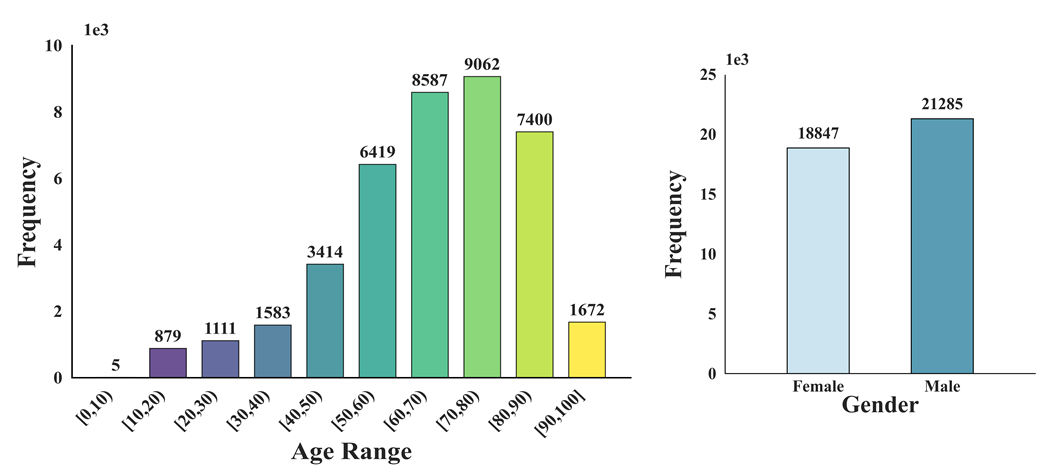}
  \caption{Patient's personal information distribution. The left figure shows the age distribution histogram, while the right figure displays the gender distribution histogram.}
  \label{fig:Patient's personal information}
\end{figure}

%\subsubsection{Data Analysis} In this section, we analyze various
\subsection{Data Analysis} In this section, we analyze various 
aspects of the MeExam dataset to better understand its characteristics.
 
\begin{figure}[t]
  \centering
  \includegraphics[scale=0.45]{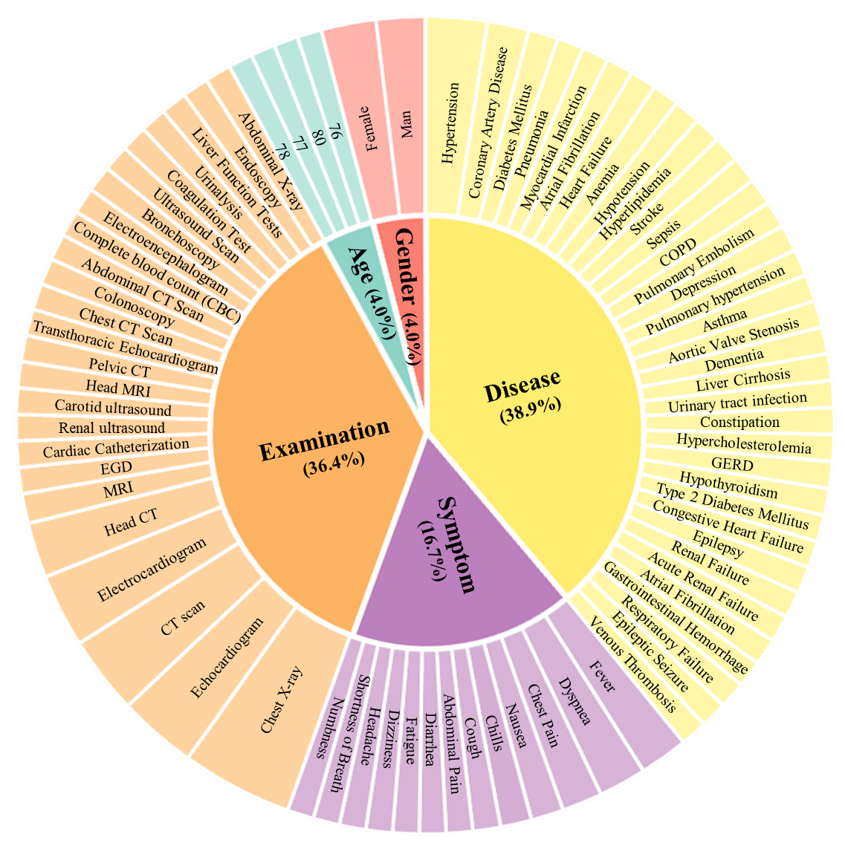}
  \caption{Entity frequency distribution of MeExam. The inner circle represents the coarse-grained entity types and the outer circle represents the fine-grained entity types, some types are denoted by abbreviations. Due to the numerous fine-grained entity types, only a subset of the types with higher entity counts in the dataset is listed here.}
  \label{fig:Entity quantity}
\end{figure}

\textbf{Data size and distribution.} The MeExam dataset contains 37,774 patient records, each with heterogeneous sequential data and demographic information. The heterogeneous sequential data includes three types of entities, namely diseases, symptoms, and examinations, while the demographic information includes age and gender. Fig~\ref{fig:Entity type} shows the distribution of fine-grained entity types, covering 1,698 diseases, 409 symptoms, and 701 examination items. Fig~\ref{fig:Patient's personal information} illustrates the age and gender distribution, with patient ages ranging from 4 to 99 years and a nearly balanced gender ratio. The age distribution indicates that old-aged individuals represent the majority, consistent with typical healthcare demographics. Additionally, Fig~\ref{fig:Entity quantity} presents the frequency distribution of entity types. Diseases are the most frequent, followed closely by examinations. Overall, the dataset comprises 956,880 entities, averaging 22.33 entities per patient, providing a rich foundation for modeling patients' health risk profiles.

\textbf{Example of MeExam Dataset}. Table~\ref{tab:Examples of Dataset} presents sample records from the MeExam dataset. In clinical practice, doctors often repeat examinations to monitor a patient's condition over time. To capture this important diagnostic behavior, we retain multiple occurrences of the same examination entities within a patient's interaction sequence. For example, patient \# 4's record includes several MRI examinations, reflecting real-world clinical patterns that are valuable for model learning.

\begin{table*}[h]

  \caption{Examples of Dataset. The sequential data and personal information of patients \#4, \#63, and \#75 are presented.}
  \label{tab:Examples of Dataset}
  \footnotesize
  \centering
%  \resizebox{\textwidth}{!}{
   \begin{tabular}{cccl}
    \toprule 
    User & Age & Gender & \multicolumn{1}{c}{Sequence Data} \\
    \midrule 
    %\multirow{2}{*}{\# 4} & \multirow{2}{*}{62} & \multirow{2}{*}
    \multirow{2}{*}{\raisebox{-0.6\height}{\# 4}} & \multirow{2}{*}{\raisebox{-0.6\height}{62}} & \multirow{2}{*}{\raisebox{-0.6\height}{female}} 
     & Rheumatoid Arthritis, Biopsy, Head CT, Headache, Nausea,Vomiting, Epilepsy, Head MRI,Diabetes Mellitus, \\
    & & & Hearing Loss, Heart Disease, MRI, MRI, MRI, Constipation, Neurological Examination, Complete blood \\
    & & &count (CBC) \\
    \midrule 
    \multirow{3}{*}{\# 63} & \multirow{3}{*}{64} & \multirow{3}{*}{man} & Melanoma, Aphasia, Epilepsy, Electroencephalogram (EEG), Sepsis, Shock, Complete blood count (CBC),\\
    & & & Cerebrospinal Fluid Analysis, Pulmonary Embolism, Hypertension, Lung cancer, Asthma, \\
    & & & Computed Tomography (CT) scan, MRI, Hemorrhage, MRI, Echocardiogram\\
    \midrule 
    \multirow{5}{*}{\# 75} & \multirow{5}{*}{61} & \multirow{5}{*}{man} & Hypertension, Prostate Cancer, Gastroesophageal reflux disease (GERD), Cerebral ischemia, Gout,\\
    & & & Chronic Obstructive Pulmonary Disease (COPD), Dyspnea, Fever, Chills, Chest X-ray,Bronchoscopy,\\
    & & & Arterial blood gas (ABG) analysis, Electrocardiogram, Electrocardiogram, Myocardial Infarction, Chest X-ray, \\
    & & & Chest X-ray, Bronchoscopy, Atrial Fibrillation, Chest CT Scan, Chest Spiral CT, \\
    & & & Pulmonary Embolism, Coagulation Test, Chest X-ray\\
    \midrule
  \end{tabular}
%}
\end{table*}

\section{Experiments}
To evaluate the effectiveness and superiority of our proposed DST-GKAN model, we conduct extensive experiments on the constructed MeExam dataset. Specifically, we aim to answer the following three Research Questions (RQ):
\begin{itemize}
    \item \textbf{RQ1:} How does DST-GKAN perform compared with the state-of-the-art recommendation baselines?
    \item \textbf{RQ2:} How do different components affect the overall performance of DST-GKAN?
    \item \textbf{RQ3:} How do different hyperparameters of DST-GKAN affect the overall recommendation performance?
\end{itemize}

\subsection{Experimental Setting}

\subsubsection{Datasets and Evaluation Metrics}
Table~\ref{tab:1 dataset} summarizes the entity and relation statistics of the MeExam dataset. We adopt the leave-one-out strategy~\cite{10598127, 10597765, 10597986} for dataset partitioning, where the last examination interaction in each patient's sequence is used for testing, and the remaining interactions form the training set. To reduce computational overhead, we follow prior work~\cite{he2017neural, zhang2019deep, 10597701, 10184783} by randomly sampling 99 negative instances for each positive instance during testing. All test samples are used for performance evaluation.

\begingroup
\renewcommand{\arraystretch}{1.5} % Default value: 1

\begin{table}[h]
    \centering
    \footnotesize
    \caption{Statistics of the collected dataset}
    \resizebox{\columnwidth}{!}{
    \begin{tabular}{cccc}
        \hline
        \textbf{Entities} & \textbf{\#Entities} & \textbf{Relations} & \textbf{\#Relations} \\ \hline
        Patients & 37,774 & Patient-Disease & 372,363 \\ 
        Diseases & 1,698 & Patient-Symptom & 160,186  \\
        Symptoms & 409 & Patient-Examination & 348,783  \\
        Examinations & 701 & - & - \\ \hline  \\
    \end{tabular}
    }
    \label{tab:1 dataset}
\end{table}
To evaluate the performance of the medical examination recommendation, we adopt two widely used metrics: HR@\(K\) and NDCG@\(K\), where \(K\) is set to 5 and 10. Hit Ratio(HR) measures whether the test examination is present in the top-\(K\) examination ranking list, while Normalized Discounted Cumulative Gain(NDCG) evaluates the location where the test examination is located, rather than only considering whether it appears. Higher HR@\(K\) and NDCG@\(K\) values reflect superior recommendation performance.

\subsubsection{Baselines and Hyperparameter Settings}
To evaluate the effectiveness of DST-GKAN, we compare it with several representative recommendation models:
\begin{itemize}
    \item \textbf{GRU4Rec} \cite{hidasi2015session} uses a GRU encoder with ranking-based loss to model user sequences for recommendations.
    \item \textbf{NARM} \cite{li2017neural} extends GRU4Rec by introducing attention mechanisms for better sequence encoding.
    \item \textbf{Caser} \cite{tang2018personalized} employs CNNs in both vertical and horizontal directions to capture sequential patterns.
    \item \textbf{SASRec} \cite{kang2018self} introduces a unidirectional Transformer to model users' evolving preferences.
    \item \textbf{BERT4Rec} \cite{sun2019bert4rec} leverages a bidirectional Transformer to learn dynamic user preferences accurately.
    \item \textbf{GCSAN} \cite{xu2019graph} integrates GNNs and self attention mechanisms to capture long-term preferences of users.
    \item \textbf{S3Rec} \cite{zhou2020s3} pre-trains on self-supervised tasks to enhance sequence modeling with contextual information.
    \item \textbf{LightSANs} \cite{fan2021lighter} introduces low-rank decomposed self-attention to efficiently capture users' latent preferences.
    \item \textbf{CORE} \cite{hou2022core} constructs session representations by linearly combining item embeddings, maintaining consistency between item and session representation spaces.
    \item \textbf{DiffRec} \cite{wang2023diffusion} employs diffusion models to generate high-quality interaction probabilities through denoising processes.
\end{itemize}

We implement the DST-GKAN model using Python 3.7 and PyTorch 1.10.0 in a non-distributed setup. All experiments are conducted on a Linux server with two NVIDIA 4090 GPUs. To ensure fair comparisons and optimal performance, we apply grid search to tune key hyperparameters. Specifically, we explore the number of patients’ rebuilt neighbors within the range [10, 20, 30, 40, 50], the number of RGAT layers within the range [1, 2, 3, 4], and the threshold of task-adaptive mechanism ranging from [0.2, 0.4, 0.6, 0.8, 1.0].

\subsection{Performance Comparison (RQ1)}
To verify the superiority and effectiveness of our proposed DST-GKAN model, we compare its recommendation performance with all the baselines. The experimental results are detailed in Table~\ref{table:2 performance}, with optimal results indicated in \textbf{bold} and suboptimal results \underline{underlined}. By analyzing the results, we can draw the following conclusions: 

\begin{table}[ht]
    \centering
    \footnotesize
    \caption{Performance comparison of different methods on the medical examination recommendation task.}
    \resizebox{\columnwidth}{!}{
    \begin{tabular}{c|cccc}
        \hline
        Models & HR@5 & HR@10 & NDCG@5 & NDCG@10 \\
        \hline
        GRU4Rec & 0.4945 & 0.6337 & 0.3568 & 0.4020 \\
        NARM & 0.5129 & 0.6510 & 0.3691 & 0.4138 \\
        Caser & 0.3731 & 0.5226 & 0.2394 & 0.2877 \\
        SASRec & 0.4845 & 0.6350 & 0.3262 & 0.3749 \\
        BERT4Rec & 0.3608 & 0.4471 & 0.2237 & 0.2516 \\
        GSCAN & 0.4509 & 0.5880 & 0.3231 & 0.3675 \\
        S3Rec & 0.4576 & 0.6032 & 0.2960 & 0.3433 \\
        LightSANs & \($\underline{0.5681}$\) & \($\underline{0.7069}$\)  & \($\underline{0.4090}$\) & \($\underline{0.4540}$\)  \\
        CORE & 0.5123 & 0.6558 & 0.3591 & 0.4056 \\
        DiffRec & 0.4689 & 0.5946 & 0.3724 & 0.4131 \\
        \textbf{Ours} & \textbf{0.5765} & \textbf{0.7302} & \textbf{0.4138} & \textbf{0.4646} \\
        % \hline
        % \% Improve. & 1.48\% & 3.30\% & 1.17\% & 2.33\% \\
        \hline
    \end{tabular}
    }
    \label{table:2 performance}
\end{table}

% \begin{table}[ht]
%     \centering
%     \caption{Performance comparison of different methods on the medical examination recommendation task.}
%     \begin{tabular}{c|cccc}
%         \hline
%         Models & HR@5 & HR@10 & NDCG@5 & NDCG@10 \\
%         \hline
%         GRU4Rec & \good{50} 0.4945 & \good{60} 0.6337 & \good{35} 0.3568 & \good{40} 0.4020 \\
%         NARM & \good{55} 0.5129 & \good{65} 0.6510 & \good{40} 0.3691 & \good{45} 0.4138 \\
%         Caser & \good{30} 0.3731 & \good{40} 0.5226 & \good{20} 0.2394 & \good{25} 0.2877 \\
%         SASRec & \good{50} 0.4845 & \good{60} 0.6350 & \good{35} 0.3262 & \good{40} 0.3749 \\
%         BERT4Rec & \good{25} 0.3608 & \good{35} 0.4471 & \good{20} 0.2237 & \good{25} 0.2516 \\
%         GSCAN & \good{45} 0.4509 & \good{55} 0.5880 & \good{35} 0.3231 & \good{40} 0.3675 \\
%         S3Rec & \good{40} 0.4576 & \good{50} 0.6032 & \good{30} 0.2960 & \good{35} 0.3433 \\
%         LightSANs & \good{80} \underline{0.5681} & \good{80} \underline{0.7069}  & \good{80} \underline{0.4090} & \good{80} \underline{0.4540}  \\
%         CORE & \good{60} 0.5123 & \good{65} 0.6558 & \good{50} 0.3591 & \good{55} 0.4056 \\
%         DiffRec & \good{55} 0.4689 & \good{60} 0.5946 & \good{55} 0.3724 & \good{60} 0.4131 \\
%         \textbf{Ours} & \good{100} \textbf{0.5765} & \good{100} \textbf{0.7302} & \good{100} \textbf{0.4138} & \good{100} \textbf{0.4646} \\
%         \hline
%         \# Improve. & 1.48\% & 3.30\% & 1.17\% & 2.33\% \\
%         \hline
%     \end{tabular}
% \end{table}
First, compared to all baseline methods, the experimental results demonstrate that DST-GKAN significantly outperforms all compared methods. The DST-GKAN’s superior performance can be attributed to four primary reasons: 1) DST-GKAN introduces a task-adaptive diffusion model to reduce the noises in heterogeneous medical data, effectively generating a recommendation-relevant subgraph for downstream examination recommendations. 2) A novel spatiotemporal graph KANsformer better extracts the temporal characteristics of patients and deeply integrates both the spatial and temporal features, thereby learning the comprehensive representations of patients for examination recommendations.

Second, compared to the sequential recommendation methods that only model the temporal information, such as GRU4Rec, Caser, NARM, SASRec, BERT4Rec, S3Rec, LightSANs, and CORE, the DST-GKAN model exhibits superior recommendation performance. These results indicate that traditional sequence encoders struggle to effectively model the heterogeneous relationships among different entities in the interaction sequences of patients. 

Finally, compared to both the transformer-based methods (GCSAN and DiffRec) and sequential methods, our proposed model achieves better overall performance, demonstrating that solely capturing features within the spatial or temporal views results in suboptimal patient and examination representations. This further highlights the necessity of incorporating spatial and temporal views in the modeling process. 

\subsection{Ablation Study (RQ2)}
\label{Ablation Study}
In this section, we conduct an ablation study to assess the impact of distinct components in our DST-GKAN model. This process involves systematically eliminating each component, while maintaining the integrity of the remaining components, resulting in three unique DST-GKAN variants. 
\begin{itemize}
    \item \textbf{w/o. cross-attention} denotes the removal of the task-adaptive denoise with the diffusion model.
    \item \textbf{w/o. RGAT} denotes the removal of the spatial RGAT encoder.
    \item \textbf{w/o. KANsformer} denotes the removal of the temporal KANsformer.
\end{itemize}
Furthermore, we assess the complete DST-GKAN model. The experimental results are presented in Table~\ref{tab:3 ablation study}, with optimal results indicated in \textbf{bold} and suboptimal results \underline{underlined}. 

\begin{table}[h]
    \centering
    \footnotesize
    \caption{Ablation study on different components for medical examination recommendation.}
    \resizebox{\columnwidth}{!}{
    \begin{tabular}{c|cccc}
        \hline
        Ablation Setting & HR@5 & HR@10 & NDCG@5 & NDCG@10 \\
        \hline
        w/o. Diffusion & 0.5739 & 0.7338 & 0.4121 & 0.4639 \\
        w/o. RGAT & 0.5697 & 0.7238 & 0.4090 & 0.4590 \\
        w/o. KANsformer & 0.5594 & 0.7160 & 0.4016 & 0.4523 \\
        \textbf{Ours} & \textbf{0.5765} & \textbf{0.7302} & \textbf{0.4148} & \textbf{0.4646} \\
        \hline
    \end{tabular}
    }
    \label{tab:3 ablation study}
\end{table}

By analyzing the results, we can obtain the following conclusions. First, removing any component invariably diminishes medical examination recommendation efficacy. %This observation emphasizes the integral role of each component within the DST-GKAN model in capturing the precise health risk status of patients. 
Second, the absence of the KANsformer (“w/o. KANsformer”) incurs a more substantial adverse impact than the removal of the RGAT (“w/o. RGAT”), further underscoring the importance of temporal modeling. Finally, when the diffusion module (“w/o. Diffusion”) is removed, there is a decline in recommendation performance.
%This finding suggests that only a subset of entities and relationships in the heterogeneous graph of patients’ interactions are truly relevant to the downstream recommendation task.

\subsection{Further Analysis (RQ2)}

To further investigate the impact of the task adaptive mechanism and cross-attention fusion mechanism, we conduct a series of comparative experiments.

\subsubsection{Effect of Task Adaptive Mechanism}
The task adaptive mechanism in the diffusion module is designed to ensure that the generated subgraph is strongly relevant to the downstream recommendation task. To confirm its effectiveness, we design a comparative experiment. The first group discards the task adaptive mechanism, which is termed as “w/o. task adaptive”. The second group retains the task adaptive mechanism, which is denoted as “w. task adaptive”. The experimental results are detailed in Figure~\ref{fig:5 task_adaptive}(a). From Figure~\ref{fig:5 task_adaptive}(a), we observe that discarding the task adaptive mechanism damages model performance, which is consistent with our expectations. This is due to the efficient identification of irrelevant data through our task adaptive mechanism on loss.

\begin{figure}[ht]
    \centering
    \includegraphics[width=\linewidth]{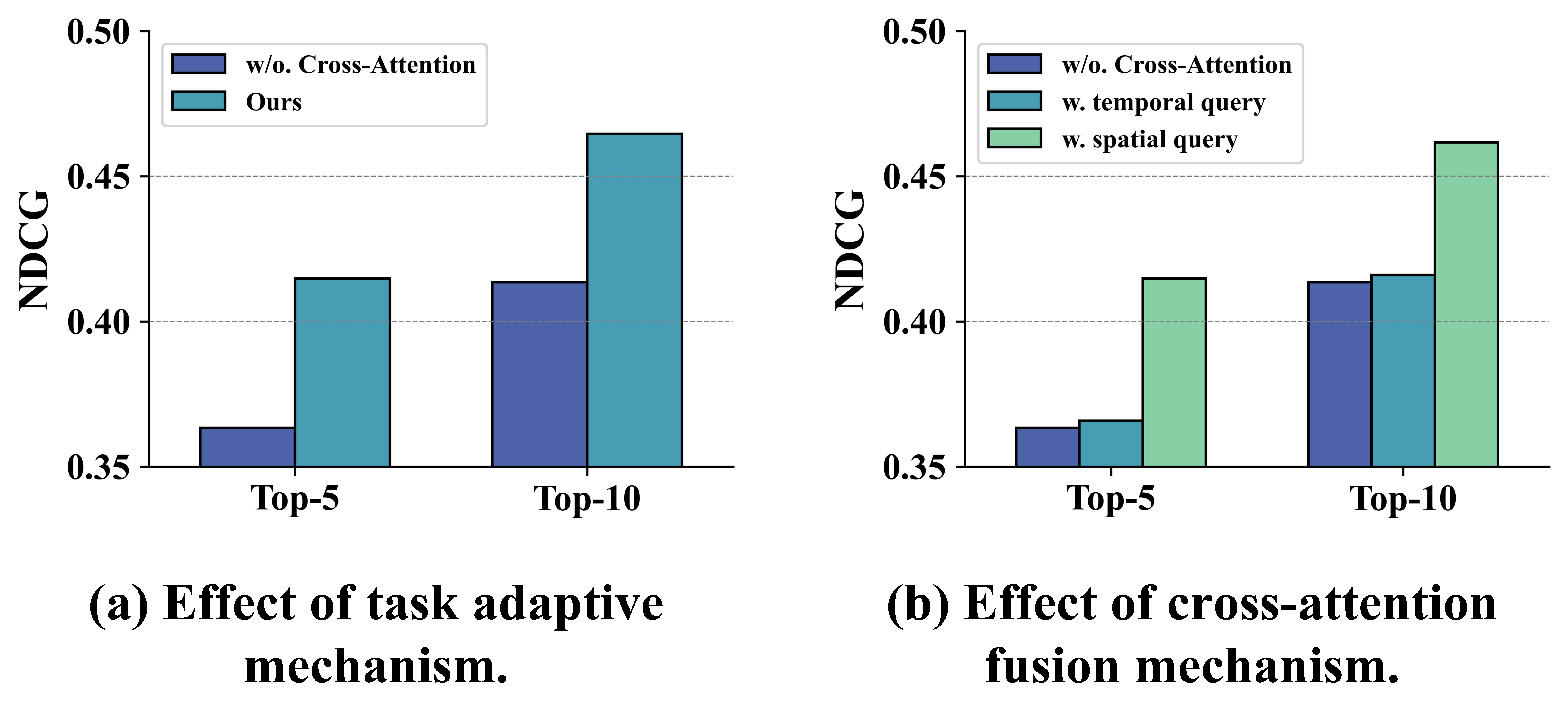}
    \caption{Effect of task adaptive mechanism and cross-attention fusion mechanism.}
    \label{fig:5 task_adaptive}
\end{figure}

% \begin{figure}[ht]
%     \centering
%     \begin{minipage}[t]{0.49\linewidth} 
%         \centering
%         \includegraphics[width=\textwidth]{Effect_of_task_adaptive_mechanism.png}
%         \label{fig:rebuilt_neighbors}
%         \caption{Effect of task adaptive mechanism.}
%     \end{minipage}
%     \hfill
%     \begin{minipage}[t]{0.49\linewidth}
%         \centering
%         \includegraphics[width=\textwidth]{Effect_of_cross_attention_fusion_mechanism.png}
%         \label{fig:rgat_layers}
%         \caption{Effect of cross-attention fusion mechanism.}
%     \end{minipage}
%     \label{fig:combined}
% \end{figure}

\subsubsection{Effect of Cross-Attention Fusion Mechanism}
%A comparative experiment is conducted to investigate the impact of cross-attention fusion mechanism, as shown in Fig~\ref{fig:5 task_adaptive}(b).
%The first group realizes the integration of spatiotemporal features using a linear layer instead of cross-attention, denoted as “w/o. cross-attention”. The second group takes local temporal features as the query and global spatial features as key and value for cross-attention fusion, denoted as “w. temporal query”. The third group takes global spatial features as the query and local temporal features as the key and value for cross-attention fusion, denoted as “w. spatial query”. %
To investigate the impact of the cross-attention fusion mechanism, we conducted a comparative experiment with different model configurations, as shown in Fig~\ref{fig:5 task_adaptive}(b).
\begin{itemize}
    \item \textbf{w/o. cross-attention} realizes the integration of spatiotemporal features using a linear layer instead of cross-attention
    \item \textbf{w/o. temporal query} takes local temporal features as the query and global spatial features as the key and value for cross-attention fusion
    \item \textbf{w/o. spatial query} takes global spatial features as the query and local temporal features as the key and value for cross-attention fusion
\end{itemize}

The experimental results lead to several crucial insights.
First, both “w. temporal query” and “w. spatial query” outperform the linear layer-based fusion method “w/o. cross-attention”, indicating that the cross-attention fusion mechanism can effectively integrate temporal and spatial features to yield superior representations of patients. Second, the “w. spatial query” method augments temporal features with spatial features as the query to generate the final patient representations. This emphasizes the importance of temporal modeling in medical examination recommendation scenarios, which is consistent with the conclusions illustrated in section~\ref{Ablation Study}. Therefore, we utilize “w/o. spatial query” for cross-attention fusion in our final model.

% \begin{figure}[ht]
%     \centering
%     \includegraphics[width=0.8\linewidth]{Effect_of_cross_attention_fusion_mechanism.png}
%     \caption{Effect of cross-attention fusion mechanism.}
%     \label{fig:cross_attention}
% \end{figure}

\subsection{Hyperparameter Sensitivity (RQ3)}

To delve into the hyperparameter sensitivity of the DST-GKAN model, our analysis concentrates on three crucial hyperparameters: the number of patients’ rebuilt neighbors, the number of RGAT layers, and the threshold $\epsilon$ of the task-adaptive mechanism.

\subsubsection{The Number of Patients’ Rebuilt Neighbors}
The number of patients’ rebuilt neighbors is a core hyperparameter of the spatial adaptive graph diffusion module, which controls the number of interactions related to the downstream recommendation task retained during the denoising process. We investigate the specific effect of the number of patients’ rebuilt neighbors within the range of [10, 20, 30, 40, 50]. As shown in Fig~\ref{fig:6 hyperparameter}(a), the optimal performance is observed when the number of patients’ rebuilt neighbors is set to 40 for our utilized dataset. Adding more neighbors might introduce additional noise, leading to reduced model accuracy.

\begin{figure}[ht]
    \centering
    \includegraphics[width=\linewidth]{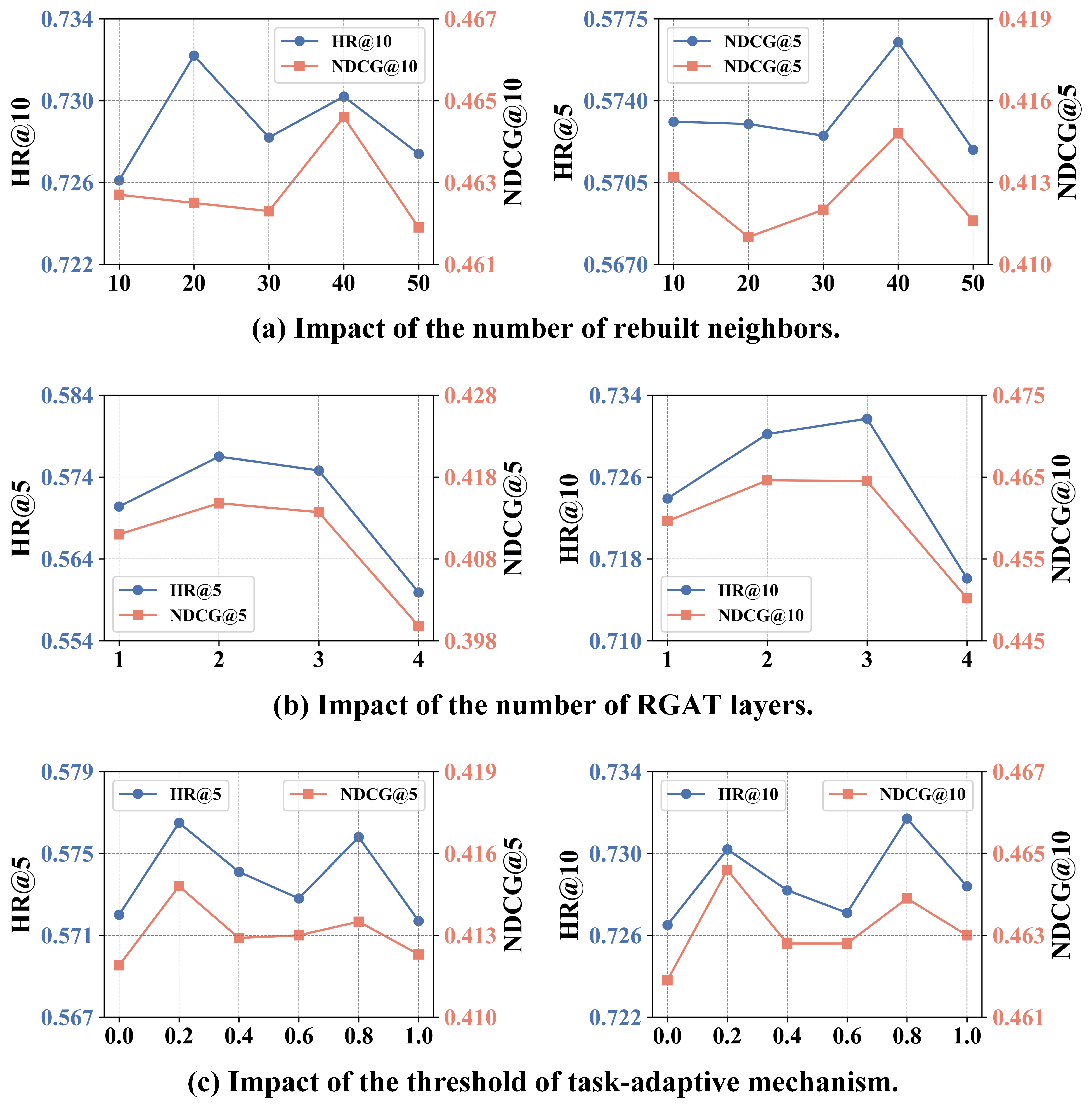}
    \caption{Impact of hyperparameter sensitivity.}
    \label{fig:6 hyperparameter}
\end{figure}

% \begin{figure}[ht]
%     \centering
%     \begin{minipage}[t]{0.49\linewidth} 
%         \centering
%         \includegraphics[width=\textwidth]{Figure 7.png}
%         \label{fig:rebuilt_neighbors}
%     \end{minipage}
%     \hfill
%     \begin{minipage}[t]{0.49\linewidth}
%         \centering
%         \includegraphics[width=\textwidth]{rebuilt10.png}
%         \label{fig:rgat_layers}
%     \end{minipage}
%     \caption{Impact of the number of patients’ rebuilt neighbors.}
%     \label{fig:combined}
% \end{figure}

\subsubsection{The Number of RGAT Layers}
The number of RGAT layers is pivotal in the DST-GKAN model, as it dictates the depth of spatial feature mining. Increasing these layers enables DST-GKAN to delve into more complex collaborative filtering signals, thereby potentially augmenting recommendation performance. We vary the number of RGAT layers from 1 to 4 while keeping the other hyperparameters constant. Fig~\ref{fig:6 hyperparameter}(b) shows that both the HR and NDCG metrics of our DST-GKAN model gradually increase as the number of RGAT layers increases. This phenomenon underscores the model’s capacity to capture high-order collaborative filtering signals from patient interactions. The optimal number of RGAT layers is 2 or 3 for the collected dataset. Conversely, overly numerous layers might integrate extraneous noise, adversely affecting the model’s accuracy.

% \begin{figure}[ht]
%     \centering
%     \begin{minipage}[t]{0.49\linewidth} 
%         \centering
%         \includegraphics[width=\textwidth]{layer_num_rgat5.png}
%         \label{fig:rebuilt_neighbors}
%     \end{minipage}
%     \hfill
%     \begin{minipage}[t]{0.49\linewidth}
%         \centering
%         \includegraphics[width=\textwidth]{layer_num_rgat10.png}
%         \label{fig:rgat_layers}
%     \end{minipage}
%     \caption{Impact of the number of RGAT layers.}
%     \label{fig:combined}
% \end{figure}

\subsubsection{The Threshold of Task-Adaptive Mechanism}
To explore the effect of the threshold of the task-adaptive mechanism on the diffusion denoising process, we vary the value of the threshold from the range of [0.0, 0.2, 0.4, 0.6, 0.8, 1.0]. As shown in Fig~\ref{fig:6 hyperparameter}(c), our proposed model achieves the optimal performance at a threshold value of 0.2. This experimental result suggests that appropriate threshold settings can effectively control the diffusion module to generate denoised heterogeneous subgraphs. These subgraphs are highly relevant to downstream tasks, thus promoting the performance of medical examination recommendations.

% \begin{figure}[ht]
%     \centering
%     \begin{minipage}[t]{0.49\linewidth} 
%         \centering
%         \includegraphics[width=\textwidth]{threshold5.png}
%         \label{fig:rebuilt_neighbors}
%     \end{minipage}
%     \hfill
%     \begin{minipage}[t]{0.49\linewidth}
%         \centering
%         \includegraphics[width=\textwidth]{threshold10.png}
%         \label{fig:rgat_layers}
%     \end{minipage}
%     \caption{Impact of the threshold of task-adaptive mechanism.}
%     \label{fig:combined}
% \end{figure}

\subsection{Case Study}

To further illustrate the reasonableness of the proposed DST-GKAN, we conduct case studies to visualize the historical interactions and recommendation results of a randomly selected user. As shown in Fig~\ref{fig:7 case study}, the selected patient 50’s previous medical records documented the following major diseases, symptoms, and examinations. In this case, the top-3 recommended list generated by the proposed model DST-GKAN includes cardiac catheterization (0.321), head CT (0.282), and neurologic examination (0.101). 

\begin{figure}[h]
    \centering
    \includegraphics[width=1\linewidth]{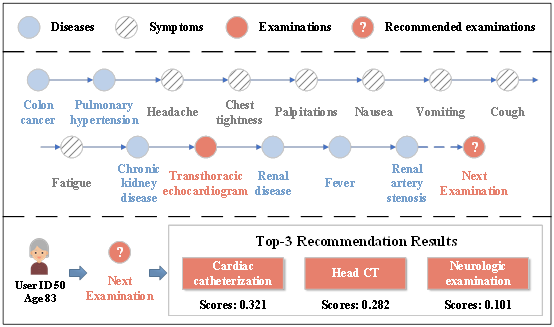}
    \caption{A case study to show the reasonableness of our proposed DST-GKAN.}
    \label{fig:7 case study}
\end{figure}

Among these, cardiac catheterization enables direct measurement of the pressure in the heart, assessment of cardiac function, and determination of pulmonary artery pressure, considering that the patient has pulmonary hypertension and symptoms of tightness and palpitations in the chest. This is important to understand the severity of pulmonary hypertension and the presence of cardiac complications. 
Second, the patient has headaches, and a CT of the head can help rule out any brain tumors, bleeding, or other lesions. This is a significant examination, especially in cancer patients, to look for any possible metastases or other neurological problems. 
Additionally, given the patient's headaches and possible neurologic symptoms, it is reasonable to perform a neurologic examination to assess neurologic function. This helps to further identify the cause of the headaches and rule out other neurologic conditions. 
In summary, the recommendation results are reasonable and consistent with the patient's history and symptoms, demonstrating that the DST-GKAN model can effectively capture the patient’s physical health status by modeling the heterogeneous information and spatiotemporal features of patients.
\section{Conclusions}
In this paper, we propose a Diffusion-driven SpatioTemporal Graph KANsformer model to first formalize the medical examination recommendation task, towards the Diagnosis assistance for patients.  
Specifically, a task-adaptive diffusion model is proposed to reduce the noises in heterogeneous medical data. And a spatiotemporal graph KANsformer is proposed to simultaneously model the complex spatial and temporal relationships. Finally, a benchmark medical examination recommendation dataset is proposed to mitigate the issues of incompleteness or lack of existing medical examination data. Extensive experiments performed on real-world medical examination datasets demonstrate the superiority of the proposed method compared to various competitive baselines.
In the future, we would like to endow the proposed DST-GKAN with the power of modeling multi-modality to
further enhance the reliability of medical examination recommendations.

\bibliographystyle{IEEEtran}
\bibliography{sample-base}

\begin{thebibliography}{10}
\providecommand{\url}[1]{#1}
\csname url@samestyle\endcsname
\providecommand{\newblock}{\relax}
\providecommand{\bibinfo}[2]{#2}
\providecommand{\BIBentrySTDinterwordspacing}{\spaceskip=0pt\relax}
\providecommand{\BIBentryALTinterwordstretchfactor}{4}
\providecommand{\BIBentryALTinterwordspacing}{\spaceskip=\fontdimen2\font plus
\BIBentryALTinterwordstretchfactor\fontdimen3\font minus \fontdimen4\font\relax}
\providecommand{\BIBforeignlanguage}[2]{{%
\expandafter\ifx\csname l@#1\endcsname\relax
\typeout{** WARNING: IEEEtran.bst: No hyphenation pattern has been}%
\typeout{** loaded for the language `#1'. Using the pattern for}%
\typeout{** the default language instead.}%
\else
\language=\csname l@#1\endcsname
\fi
#2}}
\providecommand{\BIBdecl}{\relax}
\BIBdecl

\bibitem{lopez2012property}
M.~L{\'o}pez-Nores, Y.~Blanco-Fern{\'a}ndez, J.~J. Pazos-Arias, and A.~Gil-Solla, ``Property-based collaborative filtering for health-aware recommender systems,'' \emph{Expert Systems with Applications}, vol.~39, no.~8, pp. 7451--7457, 2012.

\bibitem{shang2019gamenet}
J.~Shang, C.~Xiao, T.~Ma, H.~Li, and J.~Sun, ``Gamenet: Graph augmented memory networks for recommending medication combination,'' in \emph{proceedings of the AAAI Conference on Artificial Intelligence}, vol.~33, no.~01, 2019, pp. 1126--1133.

\bibitem{baytas2017patient}
I.~M. Baytas, C.~Xiao, X.~Zhang, F.~Wang, A.~K. Jain, and J.~Zhou, ``Patient subtyping via time-aware lstm networks,'' in \emph{Proceedings of the 23rd ACM SIGKDD international conference on knowledge discovery and data mining}, 2017, pp. 65--74.

\bibitem{choi2018mime}
E.~Choi, C.~Xiao, W.~Stewart, and J.~Sun, ``Mime: Multilevel medical embedding of electronic health records for predictive healthcare,'' \emph{Advances in neural information processing systems}, vol.~31, 2018.

\bibitem{ma2018health}
T.~Ma, C.~Xiao, and F.~Wang, ``Health-atm: A deep architecture for multifaceted patient health record representation and risk prediction,'' in \emph{Proceedings of the 2018 SIAM International Conference on Data Mining}.\hskip 1em plus 0.5em minus 0.4em\relax SIAM, 2018, pp. 261--269.

\bibitem{ho2020denoising}
J.~Ho, A.~Jain, and P.~Abbeel, ``Denoising diffusion probabilistic models,'' \emph{Advances in neural information processing systems}, vol.~33, pp. 6840--6851, 2020.

\bibitem{sohl2015deep}
J.~Sohl-Dickstein, E.~Weiss, N.~Maheswaranathan, and S.~Ganguli, ``Deep unsupervised learning using nonequilibrium thermodynamics,'' in \emph{International conference on machine learning}.\hskip 1em plus 0.5em minus 0.4em\relax PMLR, 2015, pp. 2256--2265.

\bibitem{liu2024kan}
Z.~Liu, Y.~Wang, S.~Vaidya, F.~Ruehle, J.~Halverson, M.~Solja{\v{c}}i{\'c}, T.~Y. Hou, and M.~Tegmark, ``Kan: Kolmogorov-arnold networks,'' \emph{arXiv preprint arXiv:2404.19756}, 2024.

\bibitem{su2020gate}
C.~Su, S.~Gao, and S.~Li, ``Gate: graph-attention augmented temporal neural network for medication recommendation,'' \emph{IEEE Access}, vol.~8, pp. 125\,447--125\,458, 2020.

\bibitem{Zhang2017leap}
Y.~Zhang, R.~Chen, J.~Tang, W.~F. Stewart, and J.~Sun, ``Leap: learning to prescribe effective and safe treatment combinations for multimorbidity,'' in \emph{proceedings of the 23rd ACM SIGKDD international conference on knowledge Discovery and data Mining}, 2017, pp. 1315--1324.

\bibitem{tan20224sdrug}
Y.~Tan, C.~Kong, L.~Yu, P.~Li, C.~Chen, X.~Zheng, V.~S. Hertzberg, and C.~Yang, ``4sdrug: Symptom-based set-to-set small and safe drug recommendation,'' in \emph{Proceedings of the 28th ACM SIGKDD Conference on Knowledge Discovery and Data Mining}, 2022, pp. 3970--3980.

\bibitem{liu2024leader}
Q.~Liu, X.~Wu, X.~Zhao, Y.~Zhu, Z.~Zhang, F.~Tian, and Y.~Zheng, ``Large language model distilling medication recommendation model,'' \emph{arXiv preprint arXiv:2402.02803}, 2024.

\bibitem{choi2016retain}
E.~Choi, M.~T. Bahadori, J.~Sun, J.~Kulas, A.~Schuetz, and W.~Stewart, ``Retain: An interpretable predictive model for healthcare using reverse time attention mechanism,'' \emph{Advances in neural information processing systems}, vol.~29, 2016.

\bibitem{su2022tahdnet}
Y.~Su, Y.~Shi, W.~Lee, L.~Cheng, and H.~Guo, ``Tahdnet: Time-aware hierarchical dependency network for medication recommendation,'' \emph{Journal of Biomedical Informatics}, vol. 129, p. 104069, 2022.

\bibitem{liu2022multi}
S.~Liu, X.~Wang, Y.~Xiang, H.~Xu, H.~Wang, and B.~Tang, ``Multi-channel fusion lstm for medical event prediction using ehrs,'' \emph{Journal of Biomedical Informatics}, vol. 127, p. 104011, 2022.

\bibitem{liang2024dual}
S.~Liang, X.~Li, C.~Li, Y.~Lei, Y.~Hou, and T.~Ma, ``Dual-granularity medication recommendation based on causal inference,'' \emph{arXiv preprint arXiv:2403.00880}, 2024.

\bibitem{raza2024twostage}
S.~Raza and C.~Ding, ``Improving clinical decision making with a two-stage recommender system,'' \emph{IEEE/ACM Transactions on Computational Biology and Bioinformatics}, vol.~21, no.~5, pp. 1180--1190, 2024.

\bibitem{liu2023llm}
\BIBentryALTinterwordspacing
S.~Liu, A.~P. Wright, B.~L. Patterson, J.~P. Wanderer, R.~W. Turer, S.~D. Nelson, A.~B. McCoy, D.~F. Sittig, and A.~Wright, ``{Using AI-generated suggestions from ChatGPT to optimize clinical decision support},'' \emph{Journal of the American Medical Informatics Association}, vol.~30, no.~7, pp. 1237--1245, 04 2023. [Online]. Available: \url{https://doi.org/10.1093/jamia/ocad072}
\BIBentrySTDinterwordspacing

\bibitem{wang2023tag}
\BIBentryALTinterwordspacing
Y.~Wang, S.~Ge, X.~Zhao, X.~Wu, T.~Xu, C.~Ma, and Z.~Zheng, ``Doctor specific tag recommendation for online medical record management,'' in \emph{Proceedings of the 29th ACM SIGKDD Conference on Knowledge Discovery and Data Mining}, ser. KDD '23.\hskip 1em plus 0.5em minus 0.4em\relax New York, NY, USA: Association for Computing Machinery, 2023, p. 5150–5161. [Online]. Available: \url{https://doi.org/10.1145/3580305.3599810}
\BIBentrySTDinterwordspacing

\bibitem{tian2019drgan}
B.~Tian, Y.~Zhang, X.~Chen, C.~Xing, and C.~Li, ``Drgan: A gan-based framework for doctor recommendation in chinese on-line qa communities,'' in \emph{Database Systems for Advanced Applications}, G.~Li, J.~Yang, J.~Gama, J.~Natwichai, and Y.~Tong, Eds.\hskip 1em plus 0.5em minus 0.4em\relax Cham: Springer International Publishing, 2019, pp. 444--447.

\bibitem{zheng2022ddr}
\BIBentryALTinterwordspacing
Z.~Zheng, Z.~Qiu, H.~Xiong, X.~Wu, T.~Xu, E.~Chen, and X.~Zhao, ``Ddr: Dialogue based doctor recommendation for online medical service,'' in \emph{Proceedings of the 28th ACM SIGKDD Conference on Knowledge Discovery and Data Mining}, ser. KDD '22.\hskip 1em plus 0.5em minus 0.4em\relax New York, NY, USA: Association for Computing Machinery, 2022, p. 4592–4600. [Online]. Available: \url{https://doi.org/10.1145/3534678.3539201}
\BIBentrySTDinterwordspacing

\bibitem{yin2024dataset}
M.~Yin, H.~Wang, W.~Guo, Y.~Liu, S.~Zhang, S.~Zhao, D.~Lian, and E.~Chen, ``Dataset regeneration for sequential recommendation,'' in \emph{Proceedings of the 30th ACM SIGKDD Conference on Knowledge Discovery and Data Mining}, 2024, pp. 3954--3965.

\bibitem{zhou2025dual}
\BIBentryALTinterwordspacing
Y.~Zhou, H.~Chu, Q.~Li, J.~Li, S.~Zhang, F.~Zhu, J.~Hu, L.~Wang, and W.~Yang, ``Dual-tower model with semantic perception and timespan-coupled hypergraph for next-basket recommendation,'' \emph{Neural Networks}, vol. 184, p. 107001, 2025. [Online]. Available: \url{https://www.sciencedirect.com/science/article/abs/pii/S0893608024009304}
\BIBentrySTDinterwordspacing

\bibitem{tuan20173d}
T.~X. Tuan and T.~M. Phuong, ``3d convolutional networks for session-based recommendation with content features,'' in \emph{Proceedings of the eleventh ACM conference on recommender systems}, 2017, pp. 138--146.

\bibitem{yan2019cosrec}
A.~Yan, S.~Cheng, W.-C. Kang, M.~Wan, and J.~McAuley, ``Cosrec: 2d convolutional neural networks for sequential recommendation,'' in \emph{Proceedings of the 28th ACM international conference on information and knowledge management}, 2019, pp. 2173--2176.

\bibitem{tang2018personalized}
J.~Tang and K.~Wang, ``Personalized top-n sequential recommendation via convolutional sequence embedding,'' in \emph{Proceedings of the eleventh ACM international conference on web search and data mining}, 2018, pp. 565--573.

\bibitem{yuan2019simple}
F.~Yuan, A.~Karatzoglou, I.~Arapakis, J.~M. Jose, and X.~He, ``A simple convolutional generative network for next item recommendation,'' in \emph{Proceedings of the twelfth ACM international conference on web search and data mining}, 2019, pp. 582--590.

\bibitem{hidasi2015session}
B.~Hidasi, A.~Karatzoglou, L.~Baltrunas, and D.~Tikk, ``Session-based recommendations with recurrent neural networks,'' 2016.

\bibitem{hidasi2018recurrent}
B.~Hidasi and A.~Karatzoglou, ``Recurrent neural networks with top-k gains for session-based recommendations,'' in \emph{Proceedings of the 27th ACM international conference on information and knowledge management}, 2018, pp. 843--852.

\bibitem{qin2017dual}
Y.~Qin, D.~Song, H.~Chen, W.~Cheng, G.~Jiang, and G.~Cottrell, ``A dual-stage attention-based recurrent neural network for time series prediction,'' \emph{arXiv preprint arXiv:1704.02971}, 2017.

\bibitem{li2017neural}
J.~Li, P.~Ren, Z.~Chen, Z.~Ren, T.~Lian, and J.~Ma, ``Neural attentive session-based recommendation,'' in \emph{Proceedings of the 2017 ACM on Conference on Information and Knowledge Management}, 2017, pp. 1419--1428.

\bibitem{hou2022core}
Y.~Hou, B.~Hu, Z.~Zhang, and W.~X. Zhao, ``Core: simple and effective session-based recommendation within consistent representation space,'' in \emph{Proceedings of the 45th international ACM SIGIR conference on research and development in information retrieval}, 2022, pp. 1796--1801.

\bibitem{zhou2020s3}
K.~Zhou, H.~Wang, W.~X. Zhao, Y.~Zhu, S.~Wang, F.~Zhang, Z.~Wang, and J.-R. Wen, ``S3-rec: Self-supervised learning for sequential recommendation with mutual information maximization,'' in \emph{Proceedings of the 29th ACM international conference on information \& knowledge management}, 2020, pp. 1893--1902.

\bibitem{kang2018self}
W.-C. Kang and J.~McAuley, ``Self-attentive sequential recommendation,'' in \emph{2018 IEEE international conference on data mining (ICDM)}.\hskip 1em plus 0.5em minus 0.4em\relax IEEE, 2018, pp. 197--206.

\bibitem{sun2019bert4rec}
F.~Sun, J.~Liu, J.~Wu, C.~Pei, X.~Lin, W.~Ou, and P.~Jiang, ``Bert4rec: Sequential recommendation with bidirectional encoder representations from transformer,'' in \emph{Proceedings of the 28th ACM international conference on information and knowledge management}, 2019, pp. 1441--1450.

\bibitem{xu2019graph}
C.~Xu, P.~Zhao, Y.~Liu, V.~S. Sheng, J.~Xu, F.~Zhuang, J.~Fang, and X.~Zhou, ``Graph contextualized self-attention network for session-based recommendation.'' in \emph{IJCAI}, vol.~19, 2019, pp. 3940--3946.

\bibitem{fan2021lighter}
X.~Fan, Z.~Liu, J.~Lian, W.~X. Zhao, X.~Xie, and J.-R. Wen, ``Lighter and better: low-rank decomposed self-attention networks for next-item recommendation,'' in \emph{Proceedings of the 44th international ACM SIGIR conference on research and development in information retrieval}, 2021, pp. 1733--1737.

\bibitem{wang2023diffusion}
W.~Wang, Y.~Xu, F.~Feng, X.~Lin, X.~He, and T.-S. Chua, ``Diffusion recommender model,'' in \emph{Proceedings of the 46th International ACM SIGIR Conference on Research and Development in Information Retrieval}, 2023, pp. 832--841.

\bibitem{kim2014guide}
L.~Kim, J.-A. Kim, and S.~Kim, ``A guide for the utilization of health insurance review and assessment service national patient samples,'' \emph{Epidemiology and health}, vol.~36, 2014.

\bibitem{johnson2016mimic}
A.~E. Johnson, T.~J. Pollard, L.~Shen, L.-w.~H. Lehman, M.~Feng, M.~Ghassemi, B.~Moody, P.~Szolovits, L.~Anthony~Celi, and R.~G. Mark, ``Mimic-iii, a freely accessible critical care database,'' \emph{Scientific data}, vol.~3, no.~1, pp. 1--9, 2016.

\bibitem{johnson2023mimic}
A.~E. Johnson, L.~Bulgarelli, L.~Shen, A.~Gayles, A.~Shammout, S.~Horng, T.~J. Pollard, S.~Hao, B.~Moody, B.~Gow \emph{et~al.}, ``Mimic-iv, a freely accessible electronic health record dataset,'' \emph{Scientific data}, vol.~10, no.~1, p.~1, 2023.

\bibitem{pollard2018eicu}
T.~J. Pollard, A.~E. Johnson, J.~D. Raffa, L.~A. Celi, R.~G. Mark, and O.~Badawi, ``The eicu collaborative research database, a freely available multi-center database for critical care research,'' \emph{Scientific data}, vol.~5, no.~1, pp. 1--13, 2018.

\bibitem{goldberger2000physiobank}
A.~L. Goldberger, L.~A. Amaral, L.~Glass, J.~M. Hausdorff, P.~C. Ivanov, R.~G. Mark, J.~E. Mietus, G.~B. Moody, C.-K. Peng, and H.~E. Stanley, ``Physiobank, physiotoolkit, and physionet: components of a new research resource for complex physiologic signals,'' \emph{circulation}, vol. 101, no.~23, pp. e215--e220, 2000.

\bibitem{gupta2018using}
P.~Gupta, P.~Malhotra, L.~Vig, and G.~Shroff, ``Using features from pre-trained timenet for clinical predictions.'' in \emph{KDH@ IJCAI}, 2018, pp. 38--44.

\bibitem{xu2023seqcare}
Y.~Xu, X.~Chu, K.~Yang, Z.~Wang, P.~Zou, H.~Ding, J.~Zhao, Y.~Wang, and B.~Xie, ``Seqcare: Sequential training with external medical knowledge graph for diagnosis prediction in healthcare data,'' in \emph{Proceedings of the ACM Web Conference 2023}, 2023, pp. 2819--2830.

\bibitem{jiang2024diffkg}
Y.~Jiang, Y.~Yang, L.~Xia, and C.~Huang, ``Diffkg: Knowledge graph diffusion model for recommendation,'' in \emph{Proceedings of the 17th ACM International Conference on Web Search and Data Mining}, 2024, pp. 313--321.

\bibitem{yang2023debiased}
Y.~Yang, C.~Huang, L.~Xia, C.~Huang, D.~Luo, and K.~Lin, ``Debiased contrastive learning for sequential recommendation,'' in \emph{Proceedings of the ACM web conference 2023}, 2023, pp. 1063--1073.

\bibitem{ashish2017attention}
V.~Ashish, ``Attention is all you need,'' \emph{Advances in neural information processing systems}, vol.~30, p.~I, 2017.

\bibitem{li2024multimodal}
K.~Li, L.~Xu, C.~Zhu, and K.~Zhang, ``A multimodal graph recommendation method based on cross-attention fusion,'' \emph{Mathematics}, vol.~12, no.~15, p. 2353, 2024.

\bibitem{sun2023attention}
M.~Sun, W.~Cui, Y.~Zhang, S.~Yu, X.~Liao, B.~Hu, and Y.~Li, ``Attention-rectified and texture-enhanced cross-attention transformer feature fusion network for facial expression recognition,'' \emph{IEEE Transactions on Industrial Informatics}, vol.~19, no.~12, pp. 11\,823--11\,832, 2023.

\bibitem{shen2024icafusion}
J.~Shen, Y.~Chen, Y.~Liu, X.~Zuo, H.~Fan, and W.~Yang, ``Icafusion: Iterative cross-attention guided feature fusion for multispectral object detection,'' \emph{Pattern Recognition}, vol. 145, p. 109913, 2024.

\bibitem{zheng2023casf}
J.~Zheng, H.~Liu, Y.~Feng, J.~Xu, and L.~Zhao, ``Casf-net: Cross-attention and cross-scale fusion network for medical image segmentation,'' \emph{Computer Methods and Programs in Biomedicine}, vol. 229, p. 107307, 2023.

\bibitem{chen2021crossvit}
C.-F.~R. Chen, Q.~Fan, and R.~Panda, ``Crossvit: Cross-attention multi-scale vision transformer for image classification,'' in \emph{Proceedings of the IEEE/CVF international conference on computer vision}, 2021, pp. 357--366.

\bibitem{10597701}
B.~Wu, Y.~Cheng, H.~Yuan, and Q.~Ma, ``When multi-behavior meets multi-interest: Multi-behavior sequential recommendation with multi-interest self-supervised learning,'' in \emph{2024 IEEE 40th International Conference on Data Engineering (ICDE)}, 2024, pp. 845--858.

\bibitem{10597799}
A.~Li, B.~Yang, H.~Huo, F.~K. Hussain, and G.~Xu, ``Structure- and logic-aware heterogeneous graph learning for recommendation,'' in \emph{2024 IEEE 40th International Conference on Data Engineering (ICDE)}, 2024, pp. 544--556.

\bibitem{10598015}
J.~Wu, J.~Chen, J.~Wu, W.~Shi, J.~Zhang, and X.~Wang, ``Bsl: Understanding and improving softmax loss for recommendation,'' in \emph{2024 IEEE 40th International Conference on Data Engineering (ICDE)}, 2024, pp. 816--830.

\bibitem{mao2023cross}
A.~Mao, M.~Mohri, and Y.~Zhong, ``Cross-entropy loss functions: Theoretical analysis and applications,'' in \emph{International conference on Machine learning}.\hskip 1em plus 0.5em minus 0.4em\relax PMLR, 2023, pp. 23\,803--23\,828.

\bibitem{jiang2021named}
H.~Jiang, D.~Zhang, T.~Cao, B.~Yin, and T.~Zhao, ``Named entity recognition with small strongly labeled and large weakly labeled data,'' in \emph{Annual Meeting of the Association for Computational Linguistics}, 2021.

\bibitem{alsentzer-etal-2019-publicly}
\BIBentryALTinterwordspacing
E.~Alsentzer, J.~Murphy, W.~Boag, W.-H. Weng, D.~Jin, T.~Naumann, and M.~McDermott, ``Publicly available clinical {BERT} embeddings,'' in \emph{Proceedings of the 2nd Clinical Natural Language Processing Workshop}.\hskip 1em plus 0.5em minus 0.4em\relax Minneapolis, Minnesota, USA: Association for Computational Linguistics, Jun. 2019, pp. 72--78. [Online]. Available: \url{https://www.aclweb.org/anthology/W19-1909}
\BIBentrySTDinterwordspacing

\bibitem{achiam2023gpt}
J.~Achiam, S.~Adler, S.~Agarwal, L.~Ahmad, I.~Akkaya, F.~L. Aleman, D.~Almeida, J.~Altenschmidt, S.~Altman, S.~Anadkat \emph{et~al.}, ``Gpt-4 technical report,'' \emph{arXiv preprint arXiv:2303.08774}, 2023.

\bibitem{glm2024chatglm}
T.~GLM, A.~Zeng, B.~Xu, B.~Wang, C.~Zhang, D.~Yin, D.~Rojas, G.~Feng, H.~Zhao, H.~Lai \emph{et~al.}, ``Chatglm: A family of large language models from glm-130b to glm-4 all tools,'' \emph{arXiv preprint arXiv:2406.12793}, 2024.

\bibitem{10598127}
L.~Zhang, X.~Zhou, Z.~Zeng, and Z.~Shen, ``Are id embeddings necessary? whitening pre-trained text embeddings for effective sequential recommendation,'' in \emph{2024 IEEE 40th International Conference on Data Engineering (ICDE)}, 2024, pp. 530--543.

\bibitem{10597765}
C.~Zhang, Q.~Han, R.~Chen, X.~Zhao, P.~Tang, and H.~Song, ``Ssdrec: Self-augmented sequence denoising for sequential recommendation,'' in \emph{2024 IEEE 40th International Conference on Data Engineering (ICDE)}, 2024, pp. 803--815.

\bibitem{10597986}
B.~Zheng, Y.~Hou, H.~Lu, Y.~Chen, W.~X. Zhao, M.~Chen, and J.-R. Wen, ``Adapting large language models by integrating collaborative semantics for recommendation,'' in \emph{2024 IEEE 40th International Conference on Data Engineering (ICDE)}, 2024, pp. 1435--1448.

\bibitem{he2017neural}
X.~He, L.~Liao, H.~Zhang, L.~Nie, X.~Hu, and T.-S. Chua, ``Neural collaborative filtering,'' in \emph{Proceedings of the 26th international conference on world wide web}, 2017, pp. 173--182.

\bibitem{zhang2019deep}
S.~Zhang, L.~Yao, A.~Sun, and Y.~Tay, ``Deep learning based recommender system: A survey and new perspectives,'' \emph{ACM computing surveys (CSUR)}, vol.~52, no.~1, pp. 1--38, 2019.

\bibitem{10184783}
Z.~Wang, X.~Chen, R.~Zhou, Q.~Dai, Z.~Dong, and J.-R. Wen, ``Sequential recommendation with user causal behavior discovery,'' in \emph{2023 IEEE 39th International Conference on Data Engineering (ICDE)}, 2023, pp. 28--40.

\end{thebibliography}

% \newpage
\vskip -2\baselineskip plus -1fil
\begin{IEEEbiography}[{\includegraphics[width=1in,height=1.25in,clip,keepaspectratio]{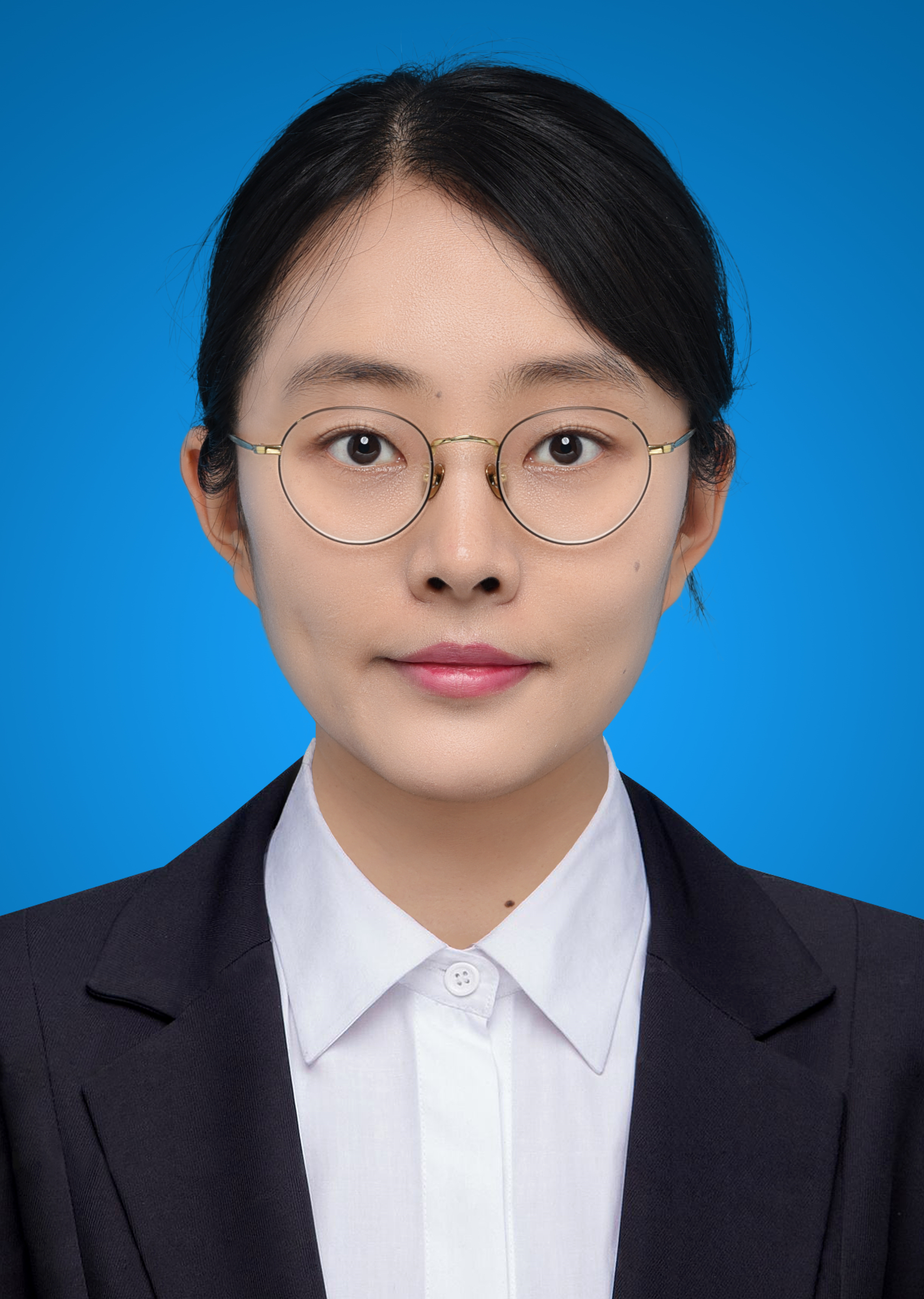}}]{Jianan Li}
is a Lecturer in the School of Computer Science and Technology at Xidian University, Xi’an, China. She received her Ph.D. degree in School of Artificial Intelligence from Xidian University in 2020. Her research interests are deep learning, recommendation system, and action recognition.
\end{IEEEbiography}
\vskip -2\baselineskip plus -1fil
\begin{IEEEbiography}[{\includegraphics[width=1in,height=1.25in,clip,keepaspectratio]{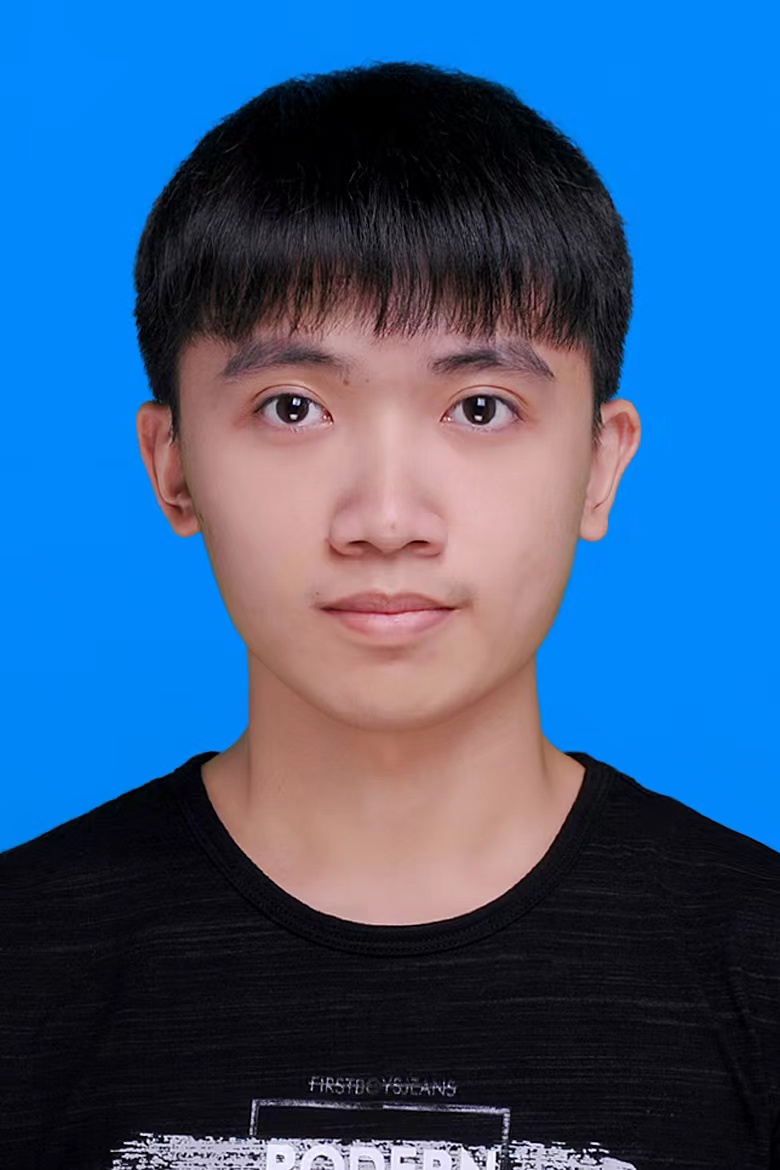}}]{Yangtao Zhou}
received his B.S. degree in the School of Computer Science and Technology at Xidian University, China, in 2020. He is currently studying as a doctoral student in software engineering at Xidian University, China. His research interests include recommendation system, knowledge graph, and data mining.
\end{IEEEbiography}
\vskip -2\baselineskip plus -1fil
\begin{IEEEbiography}[{\includegraphics[width=1in,height=1.25in,clip,keepaspectratio]{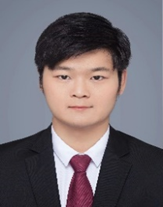}}]{Zhifu Zhao}
is a Associate Professor in the School of Artificial Intelligence at Xidian University, Xi’an, China. He received his Ph.D. degree in School of Artificial Intelligence from Xidian University in 2020. His research interests are deep learning, video understanding and compressive sensing.
\end{IEEEbiography}
\vskip -2\baselineskip plus -1fil
\begin{IEEEbiography}[{\includegraphics[width=1in,height=1.25in,clip,keepaspectratio]{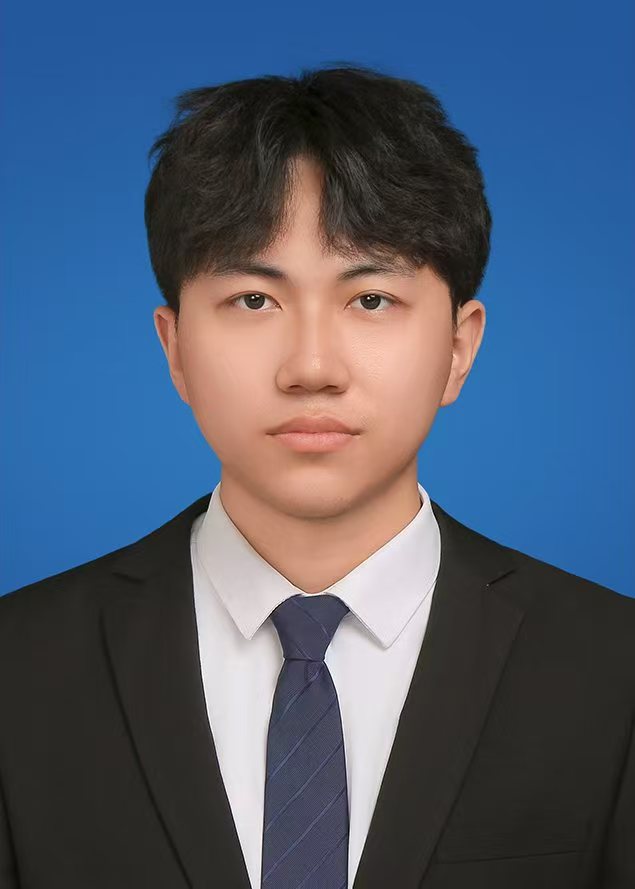}}]{Qinglan Huang}
received his B.S. degree from the School of Electrical Engineering and Information at Southwest Petroleum University, China, in 2021. He is currently pursuing a Master's degree in computer technology at Xidian University, China. His research interests include recommendation systems, knowledge graphs, and data mining.
\end{IEEEbiography}
\vskip -2\baselineskip plus -1fil
\begin{IEEEbiography}[{\includegraphics[width=1in,height=1.25in,clip,keepaspectratio]{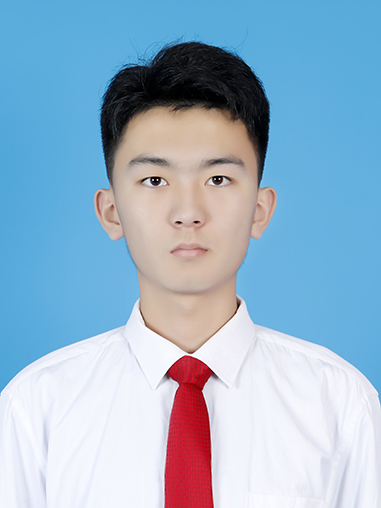}}]{Jian Qi}
is an undergraduate student at the School of Computer Science and Technology, Xidian University, China. His research interests include deep learning and recommendation systems.
\end{IEEEbiography}
\vskip -2\baselineskip plus -1fil
\begin{IEEEbiography}[{\includegraphics[width=1in,height=1.25in,clip,keepaspectratio]{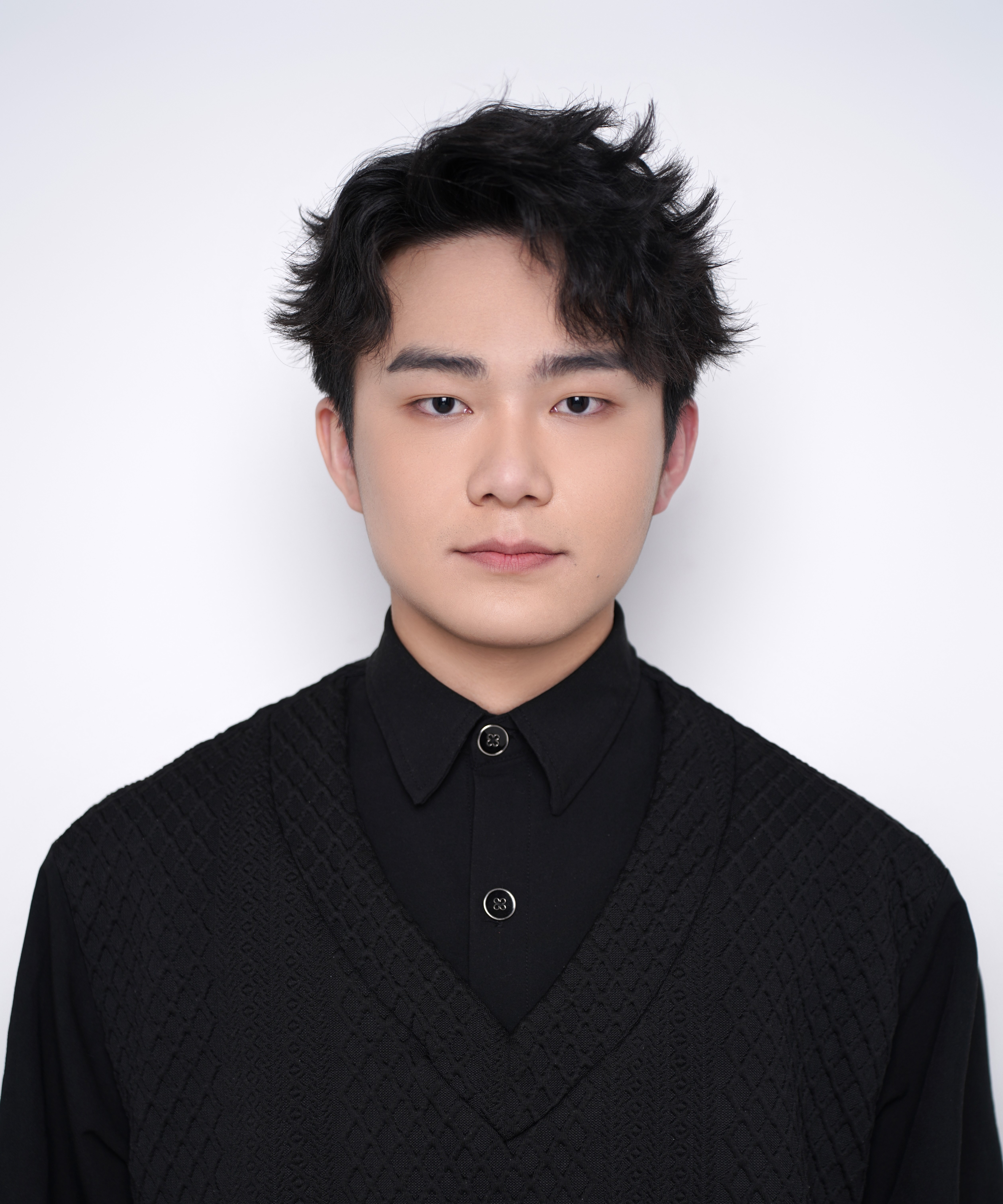}}]{Xiao He}
is an undergraduate student in the Software Engineering program at the School of Computer Science and Technology, Xidian University, Xi'an, China. He is passionate about Deep Learning. His current focus includes recommendation system.
\end{IEEEbiography}
\vskip -2\baselineskip plus -1fil
\begin{IEEEbiography}[{\includegraphics[width=1in,height=1.25in,clip,keepaspectratio]{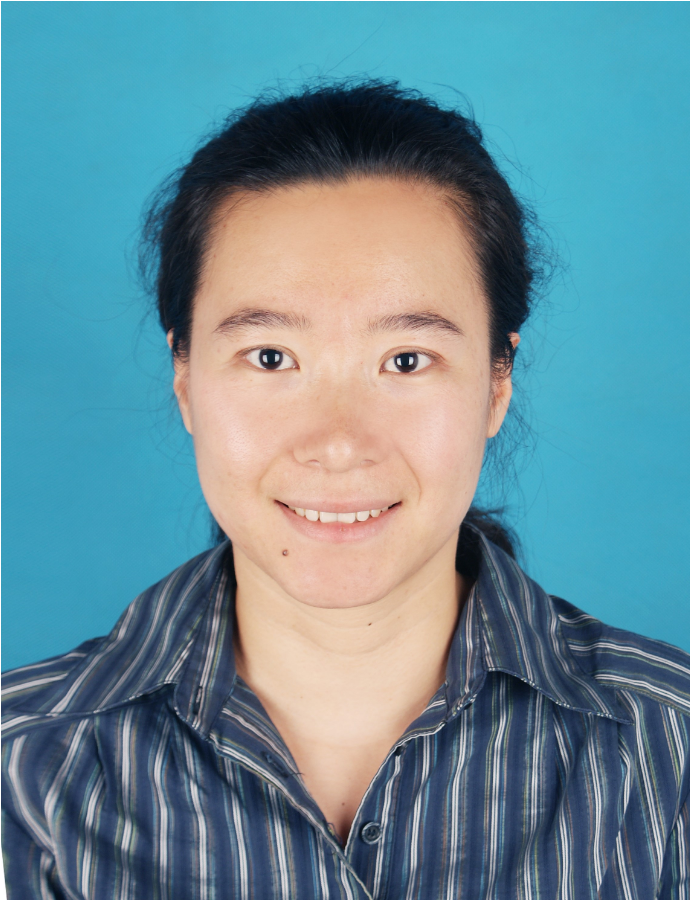}}]{Hua Chu}
received her B.S. degree in Computer and Applications and her Ph.D. degree in Computer Application Technology from Xidian University, China, in 1998 and 2007, respectively. She is currently a Professor in the School of Computer Science and Technology at Xidian University, China. Her research interests include recommendation system and object-oriented programming.
\end{IEEEbiography}
\vskip -2\baselineskip plus -1fil
\begin{IEEEbiography}[{\includegraphics[width=1in,height=1.25in,clip,keepaspectratio]{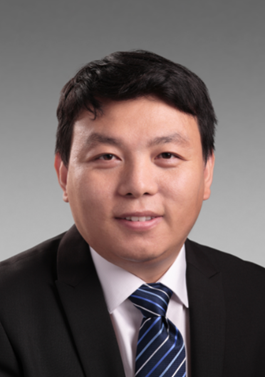}}]{Fu Li}
is a Professor in the School of Artificial Intelligence at Xidian University, Xi’an, China. He is the head of Xidian-Xilinx Embedded Digital Integrated System Joint Laboratory. He received his B.S. degree in Electronic Engineering from Xidian University in 2004, and Ph.D. degree in Electrical \& Electronic Engineering from the Xidian University in 2010. He has published more than 30 papers international and national journals, and international conferences. His research interests are brain-computer interface, deep learning, small target detection, 3D imaging, embedded deep learning, image and video compression processing, VLSI circuit design, target tracking, neural network acceleration and Implementation of intelligent signal processing algorithms (DSP \& FPGA).
\end{IEEEbiography}

%\bibitem{ref1}
%{\it{Mathematics Into Type}}. American Mathematical Society. [Online]. Available: https://www.ams.org/arc/styleguide/mit-2.pdf

%\bibitem{ref2}
%T. W. Chaundy, P. R. Barrett and C. Batey, {\it{The Printing of Mathematics}}. London, U.K., Oxford Univ. Press, 1954.

%\bibitem{ref3}
%F. Mittelbach and M. Goossens, {\it{The \LaTeX Companion}}, 2nd ed. Boston, MA, USA: Pearson, 2004.

%\bibitem{ref4}
%G. Gr\"atzer, {\it{More Math Into LaTeX}}, New York, NY, USA: Springer, 2007.

%\bibitem{ref5}M. Letourneau and J. W. Sharp, {\it{AMS-StyleGuide-online.pdf,}} American Mathematical Society, Providence, RI, USA, [Online]. Available: http://www.ams.org/arc/styleguide/index.html

%\bibitem{ref6}
%H. Sira-Ramirez, ``On the sliding mode control of nonlinear systems,'' \textit{Syst. Control Lett.}, vol. 19, pp. 303--312, 1992.

%\bibitem{ref7}
%A. Levant, ``Exact differentiation of signals with unbounded higher derivatives,''  in \textit{Proc. 45th IEEE Conf. Decis.
%Control}, San Diego, CA, USA, 2006, pp. 5585--5590. DOI: 10.1109/CDC.2006.377165.

%\bibitem{ref8}
%M. Fliess, C. Join, and H. Sira-Ramirez, ``Non-linear estimation is easy,'' \textit{Int. J. Model., Ident. Control}, vol. 4, no. 1, pp. 12--27, 2008.

%\bibitem{ref9}
%R. Ortega, A. Astolfi, G. Bastin, and H. Rodriguez, ``Stabilization of food-chain systems using a port-controlled Hamiltonian description,'' in \textit{Proc. Amer. Control Conf.}, Chicago, IL, USA,
%2000, pp. 2245--2249.

%\end{thebibliography}

\newpage

 %needed around the contents of the optional argument to biography to prevent
 %$\backslash${\tt{includegraphics}} command within an optional argument. (You can create
% your own custom macro containing the $\backslash${\tt{includegraphics}} command to make things
 

%\bf{If you include a photo:}\vspace{-33pt}
%\begin{IEEEbiography}[{\includegraphics[width=1in,height=1.25in,clip,keepaspectratio]{fig1}}]{Michael Shell}
%Use $\backslash${\tt{begin\{IEEEbiography\}}} and then for the 1st argument use $\backslash${\tt{includegraphics}} to declare and link the author photo.
%Use the author name as the 3rd argument followed by the biography text.
%\end{IEEEbiography}



%\vfill

\end{document}